\documentclass{JHEP3}
\usepackage{epsfig}
\usepackage{graphicx}
\usepackage{dcolumn}
\usepackage{amsmath}
\usepackage{enumerate}

\newcommand{\beq}{\begin{equation}}
\newcommand{\eeq}{\end{equation}}

 \def\gtap{\mathrel{ \rlap{\raise 0.511ex \hbox{$>$}}{\lower 0.511ex
   \hbox{$\sim$}}}} 
\def\ltap{\mathrel{ \rlap{\raise 0.511ex
    \hbox{$<$}}{\lower 0.511ex \hbox{$\sim$}}}} 
    
\title{The dark side of curvature}

\author{Gabriela  Barenboim \\ Departament de F\'{\i}sica Te\`orica,Universitat de
Val\`encia and IFIC\\Carrer Dr. Moliner 50, E-46100 Burjassot (Val\`encia), Spain\\
\email{Gabriela.Barenboim@uv.es}}

\author{Enrique Fern\'andez Mart\'{\i}nez\\ 
Max-Planck-Institut f\"ur Physik (Werner-Heisenberg-Institut)\\
F\"ohringer Ring 6, 80805 M\"unchen, Germany\\
\email{enfmarti@mppmu.mpg.de}}
\author{Olga Mena\\Instituto de F\'{\i}sica Corpuscular, IFIC, CSIC and Universidad de Valencia, Spain\\ \email{omena@fnal.gov}}
\author{Licia Verde \\
ICREA \& Instituto de Ciencias del Cosmos, Universitat de Barcelona, Marti i Franques 1, 08028, Barcelona, Spain  \email{liciaverde@icc.ub.edu}}

\abstract{Geometrical tests such as the combination of the  Hubble parameter $H(z)$ and
  the angular diameter distance $d_A(z)$  can, in principle, break the degeneracy between the dark energy equation of state parameter $w(z)$, and the spatial curvature $\Omega_k$ in a direct, model-independent way. In practice, constraints on these quantities  achievable from realistic experiments, such as those to be provided by  Baryon Acoustic Oscillation (BAO) galaxy surveys in combination with CMB data, can resolve  the cosmic confusion between the dark energy  equation of state parameter and curvature only statistically and within a parameterized model for $w(z)$.
 Combining measurements of both $H(z)$ and $d_{A}(z)$ up to  sufficiently high redshifts $z\sim 2$ and employing a parameterization of the redshift evolution of the dark energy equation of state are the keys to resolve the 
  $w(z)-\Omega_k$ degeneracy. }


\begin{document}

\section{Introduction}

Current cosmological measurements point to a flat universe whose mass-energy includes $5\%$ ordinary matter and $22\%$ non-baryonic dark matter, but is dominated by  a dark energy component, identified as the engine for accelerated expansion e.g.,~\cite{SpergelWMAP03,SpergelWMAP06,wmap5dunkley,wmap5komatsu,Wood-VaseySN07, TegmarkLRGDR4,PercivalLRG, BethLRG}. The present-day accelerated expansion reveals new physics missing from our universe's picture, and it constitutes the fundamental key to understand the fate of the universe.

 The most economical description of the cosmological measurements attributes the dark energy to a Cosmological Constant in Einstein's equations, representing an invariable vacuum energy density. The equation of state parameter of the dark energy component in the cosmological constant case is constant, $w=\rho/p=-1$. A dynamical option is to suppose that a cosmic scalar field $\phi$, called quintessence, changing with time and varying across space, is slowly approaching its ground state. In the quintessence scenario the equation of state is given by $w=(\dot{\phi}^2/2 -V(\phi))/(\dot{\phi}^2/2 +V(\phi))$ and in general it is not constant through cosmic time~\cite{Caldwell98,Quint99,Wang00,RatraPeebles88a,RatraPeebles88b,Wetterich95}. Another alternative is that the dark energy is an extra cosmic fluid  with a complex (time dependent) equation of state parameter $w(z)$. 

However, given the fact that we only know of dark energy from its gravitational effects, what we are trying to explain by the addition of exotic fluids could just simply be explained by corrections to Einstein gravity e.g.,~\cite{DGP,DDG02,Carroll03,Carroll05}. Although this requires a modification of Einstein's equations of gravity on large scales, this is not unexpected for an effective 4-dimensional description of higher dimensional theories. Most of the modifications of gravity proposed so far are based either on models with extra spatial dimensions or on models with an action which is non linear in the curvature scalar (that is, higher derivative theories, scalar-tensor theories or generalized functions of the Ricci scalar). 

Determining the nature of dark energy is among the major aims of future galaxy
surveys. A mandatory first step is  to extract  as precisely as possible  the
dark energy equation of state and its time dependence. Current
cosmological limits on the equation of state parameter are model-dependent. In particular, most of the reported limits rely on the assumption of an underlying
spatially flat, $\Omega_k=0$ universe. 

In this paper we focus on the cosmic
confusion between the equation of state parameter $w$ and a non-negligible spatial curvature $\Omega_k$, exploring both constant $w$
and redshift dependent $w(z)$ cases. Namely, it is quite possible that 
the universe we live in
could be of the $\Lambda $CDM-type  with a small curvature component. In such a
universe, if the curvature is assumed to be zero, one would reconstruct a $
w\neq -1$. Thus, by combining data at different redshifts, the 
  equation of state reconstructed under the incorrect assumption of zero curvature, could be a time dependent
$w(z)$. The authors of ~\cite{prev1} study the 
degeneracy between $\Omega_k$ and $w$, for constant $w$. By exploiting
luminosity distance data at different redshifts, they identify a
critical redshift $z_{cr}$(which turns out to be $\sim 3$) at which the the luminosity distance
becomes insensitive to curvature and the error on $w$ is minimal (see
also Ref.~\cite{prev4}). They conclude that the degeneracy between
$\Omega_k$ and $w$ could be alleviated if one combines luminosity
distance data at redshifts below and above $z_{cr}$, since the
$\Omega_k-w$ degeneracy at $z<z_{cr}$ is opposite to that at
$z>z_{cr}$. The authors of ~\cite{prev2} extended the previous
analysis, considering dynamical dark energy models $w(z)$, and other
fundamental observables, such as the Alcock-Pazynski
test~\cite{alcock-pazynski}. 
More recently, it has been shown~\cite{prev3} that the $w(z)$ reconstructed  assuming zero curvature from Hubble parameter $H(z)$  will have a divergence if the curvature is
negative; conversely if the curvature is positive it is the $w(z)$ reconstructed  from  the angular diameter distance $d_{A}(z)$ that will have a divergence.
The redshift position of the divergence depends on the size of $\Omega_k$.
Thus, in principle, the different behaviours for these 
two reconstructed $w(z)$'s could be used to infer both the sign and the size of the
cosmic curvature and the dynamical character of the dark energy
component. 

 Here we show that, when realistic errors on $H(z)$ and $d_{A}(z)$
expected from future BAO surveys are considered,  this is not possible: the expected signal is  smaller than the errors.
 It is still possible to separate the effects of curvature from those of dark energy, but it must be done statistically, within a  parameterized model for $w(z)$. We also forecast, using the Fisher matrix
formalism, the errors on $w(z)$ (parameterized by a  popular 2-parameter model) and
$\Omega_{k}$ using measurements from a variety of surveys with characteristics not too dissimilar from those of planned  spectroscopic and photometric galaxy surveys, in combination with forecasted constraints from a CMB experiment with characteristics similar to those of the Planck mission.
 We quantify the benefits of increased volumes and, 
in the case of photometric surveys, reduced photo-z errors. 
We show that in all these cases, the $w(z)$-$\Omega_{k}$ degeneracy is  greatly alleviated if the BAO surveys cover a redshift range up to z$\sim 2$. 

The structure of the paper is as follows. In Sec.~\ref{sec:seci} we
present the constraints on $w(z)$ and $\Omega_{k}$ from current
available data. We present the reconstructed $w(z)$ from $H(z)$ and
$d_{A}(z)$ mock data and errors in Sec.~\ref{sec:secii}. 
Section~\ref{sec:seciii} is devoted to the future $\Omega_{k}$ and $w(z)$
constraints from a variety of surveys which will cover different volumes 
and different redshift ranges. These surveys could provide the ideal 
tool to pin down both the cosmic curvature and measure the dark energy
 simultaneously. We conclude in \ref{sec:seciiii}.

\section{Current Cosmological Constraints on $\Omega_k$ and $w(z)$}
\label{sec:seci}
In this section we explore the current constraints on both a constant
and a two-parameter model for the dynamical dark energy equation of state $w(z)$, assuming a
non zero spatial curvature (see also Ref.~\cite{ichikawa,Zhao,pia}).
We work in the framework of a cosmological model described by nine free parameters~\footnote{We use the full set of parameters with both $w_0$ and $w_a$ when considering a varying $w(z)$, and a reduced set with $w=w_0$ and $w_a=0$ when considering the constant $w$ case.},
\begin{equation}
\theta=\left\{w_{b}, w_{dm}, \theta_{CMB}, \tau, \Omega_k, n_s, w_0, w_a, A_s \right\}\,,
\end{equation}
being $w_{b}=\Omega_{b} h^2$ and $w_{dm}=\Omega_{dm} h^2$ the physical baryon and dark matter densities respectively~\footnote{The current value of the Hubble parameter $H_0$ is defined as $100 h$.}, $\theta_{CMB}$  
~\footnote{The $\theta_{CMB}$ parameter can be replaced by the $H_0$ parameter. However, using $\theta_{CMB}$ is a superior choice due to its smaller correlation with the remaining parameters.} a parameter proportional to the ratio of the sound horizon to the angular diameter distance, $\tau$ the reionisation optical depth, $\Omega_k$ the spatial curvature, $n_s$ the scalar spectral index and $A_s$ the scalar amplitude. The parameterization of the dark energy equation of state we use here, in terms of the scale factor $a$, reads 
\begin{equation}     
w(a) = w_0 + w_a \ (1-a)\,,
\label{eq:eosp}
\end{equation}     
 which  has been extensively explored in the literature~\cite{chevpol,linder03,detf,linder06}. In terms of the redshift, Eq.(\ref{eq:eosp}) reads
\begin{equation}     
w(z) = w_0 + w_a \ \frac{z}{1+z}~.
\label{eq:eosp2}
\end{equation}   
We chose this parameterization because, in the absence of observational indications that $w(z)$ is not constant, it has become the standard one to use by all authors to be able to compare constraints obtained by different analysis, using different data. Of course, should any indication of a redshift-dependent $w(z)$ arise, the type of parameterization assumed would need to be a realistic fit to the data for the results to be meaningful. For the numerical simulations presented in this section we will assume
the priors $-2<w_0<0$ and $-1<w_a<1$. In this work we use the publicly
available package \texttt{cosmomc}~\cite{lewis}. The code has been modified~\cite{lewishu} for the time dependent $w(z)$ case. 
\begin{figure}[t]
\vspace{-0.1cm}
\begin{center}
\begin{tabular}{cc}
\hspace{-0.55cm} \includegraphics[width=7.5cm]{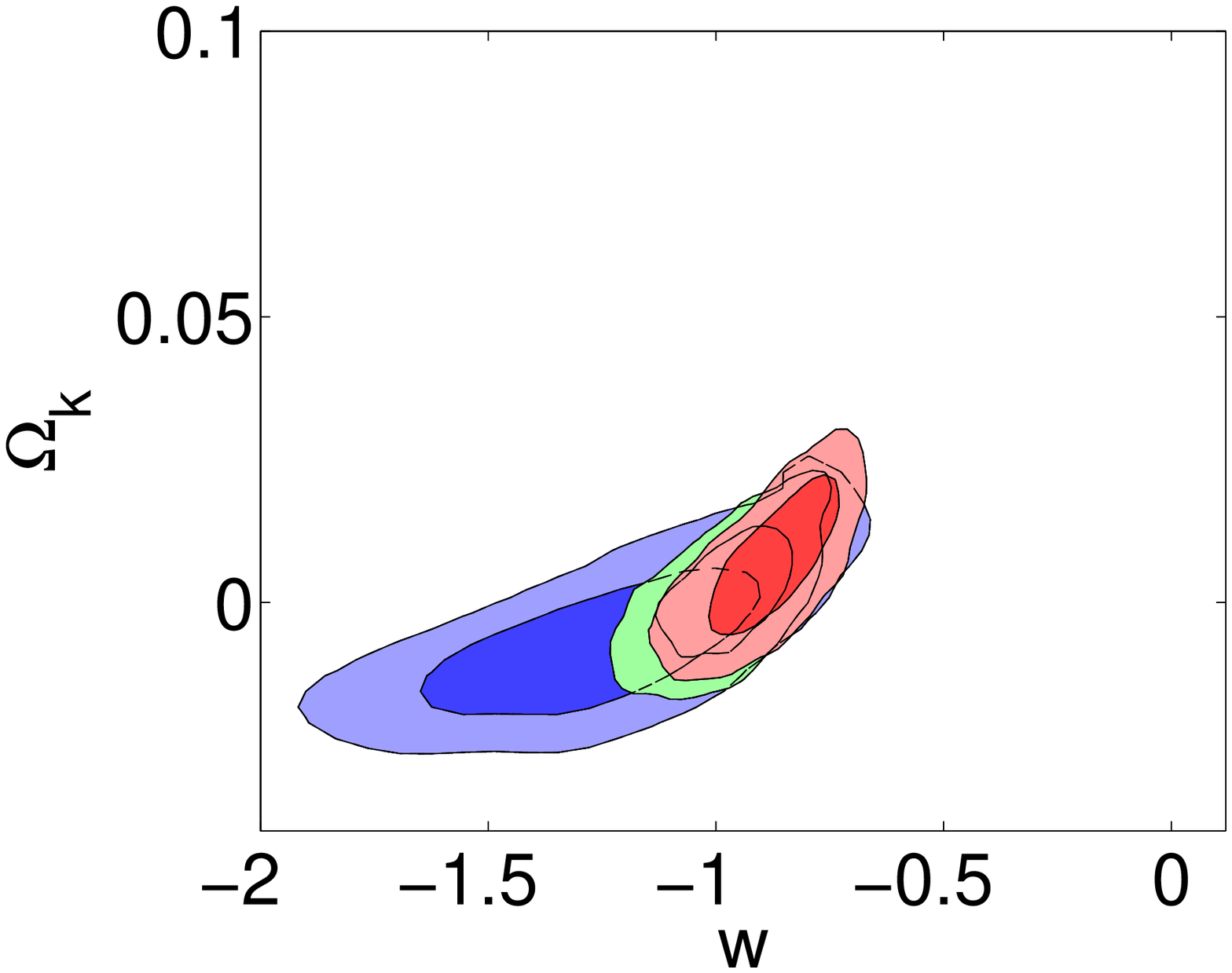} &
		 \includegraphics[width=7.5cm]{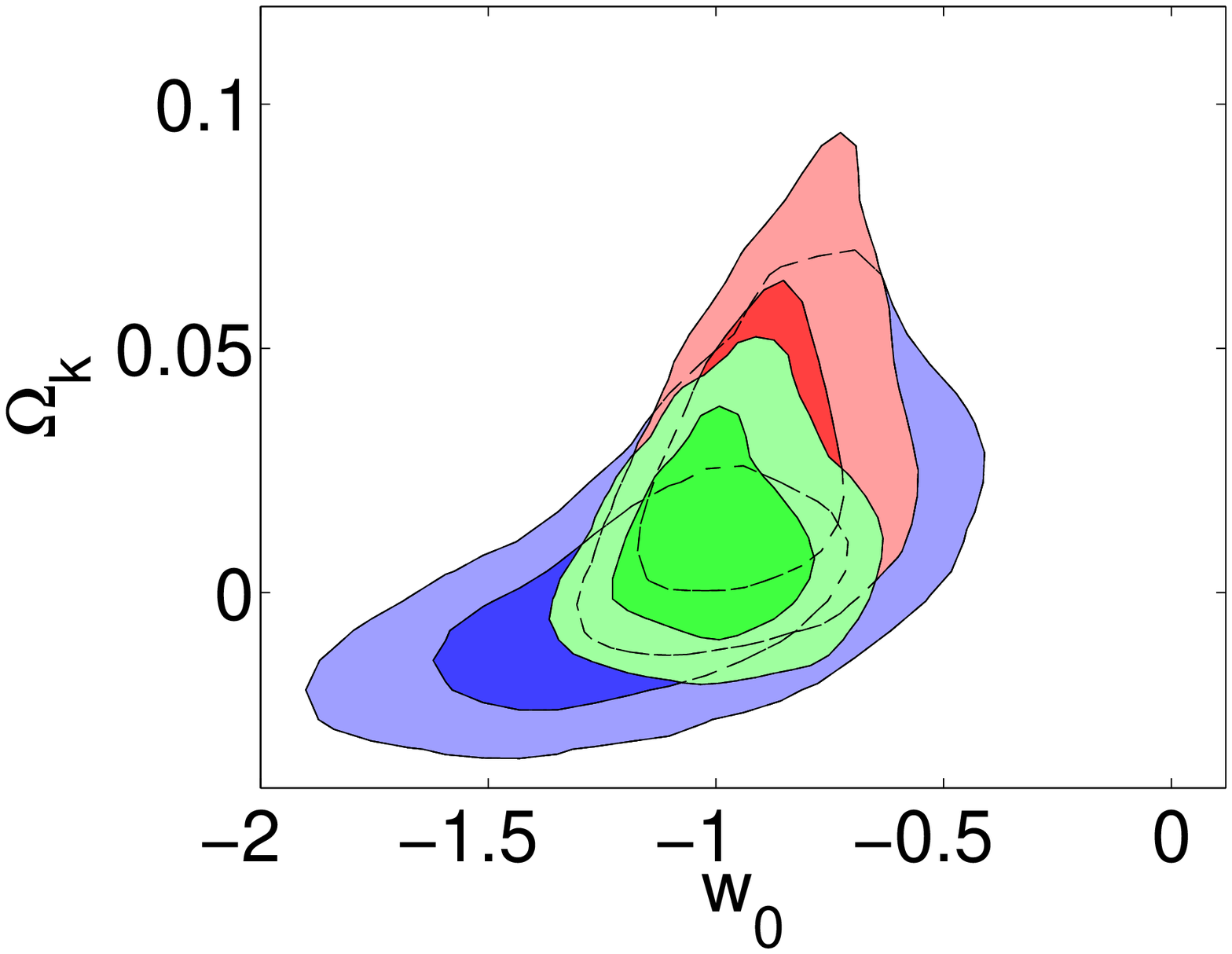}
\end{tabular}
\caption{\it Left panel: $1$ and $2-\sigma$ constraints on the $w$-$\Omega_k$ plane for a constant $w$ model. The blue, green and red contours show the $1-$ and $2-$ $\sigma$ joint confidence regions (marginalised over all other parameters) for  run0 and  runI  both combined with SDSS BAO data, and  for runII results, respectively. Right panel: $1$ and $2\sigma$ constraints on $w_0$-$\Omega_k$ plane for the full 9-parameters model where $w(z)$ is parameterized by $w_0$ and $w_a$ as in Eq. 2.2.}
\label{fig:fig0}
\end{center}
\end{figure}

On the data side, we start with a conservative compendium of
cosmological datasets. First, in what we call \emph{run0}, we include
WMAP 5-year data~\cite{wmap5dunkley,wmap5komatsu} and a prior on the
Hubble parameter of $H_0=74.2\pm3.6$~km/s/Mpc from Ref.~\cite{riessr}. We then add in \emph{runI} the constraints coming from the latest compilation of supernovae (SN) from Ref.~\cite{supernovae}.  Finally, we use the data on the matter power spectrum LSS from the spectroscopic survey of Luminous Red Galaxies (LRGs) from the Sloan Digital Sky Survey (SDSS) survey~\cite{TegmarkLRGDR4} which we refer to as the LSS data (\emph{runII}). 
In summary, 
\begin{itemize}
\item \emph{run0}=WMAP(5yr)+$H_0$
\item \emph{runI}=\emph{run0}+SN 
\item \emph{runII}= \emph{runI}+LSS
\end{itemize} 
We combine both \emph{run0} and \emph{runI} results with SDSS BAO data~\cite{sdssbao} at $z=0.35$. We first consider the case of  a constant equation of state $w$. Figure~\ref{fig:fig0} (left panel) shows the $1$ and $2\sigma$ joint constraints in the $w$-$\Omega_k$ plane from current data (marginalised over all other 6 parameters). We show the results from the three runs described above. Notice that a degeneracy is present, being $w$ and $\Omega_k$ positively correlated.
The shape of the contours can be easily understood. In a universe with a dark energy component with a $w>-1$ the distance to the last scattering surface will be shorter, effect which can be compensated in an open universe with $\Omega_{k}>0$. The opposite happens if $w<-1$. A similar analysis to the one shown in Fig.~\ref{fig:fig0} (left panel) is presented in Ref.~\cite{wmap5komatsu}. Here we use a prior on $\theta_{CMB}$ rather than on $H_0$ (as done in Ref.~\cite{wmap5komatsu}). Notice however that the tendency of the degeneracy in the $w-\Omega_k$ plane is the same in both studies.

 Next, we explore  the case of a time dependent $w(z)$. The
 parameterization we use is given by Eq.~(\ref{eq:eosp2}). The right
 panel of Fig.~\ref{fig:fig0} shows the constraints in the
 $w_{0}$-$\Omega_k$ plane. Notice that the constraints on both the
 spatial curvature $\Omega_k$ and $w_0$ are much weaker than those
 obtained when we assume a constant $w$, allowing for much larger
 positive values of $\Omega_k$ (which correspond to an open
 universe). We obtain as well less stringent constraints on $w_0$ than
 those obtained on $w$ due to the addition of the extra $w_a$
 parameter. If one relaxes the assumption of constant 
   equation of state, the $2\sigma$ marginalized error on $\Omega_{k}$
   is $\sim 0.03$ for \emph{runI} plus SSDS BAO data and $0.04$ for \emph{runII}. Comparing to the constant $w$ case this corresponds to an increase of  a factor  $\sim 1.8$  and $2.3$ respectively in the errors on $\Omega_k$.
 Notice that, even after combining with BAO data,  if $\Omega_k$ is
 positive (open universe), for a model with time dependent $w(z)$, the
 maximum allowed cosmic curvature contribution is double of that
 allowed  for a model with a constant $w$. The contours for the
 constant $w$ and non constant $w$ cases are very similar in the
 $\Omega_k<0$ region, since $\Omega_k$ can not be arbitrarily small
 (the total  energy density parameter including the curvature
 contribution needs to be 1).  
 
\begin{figure}[t]
\vspace{-0.1cm}
\begin{center}
\begin{tabular}{cc}
\hspace{-0.55cm} \includegraphics[width=7.5cm]{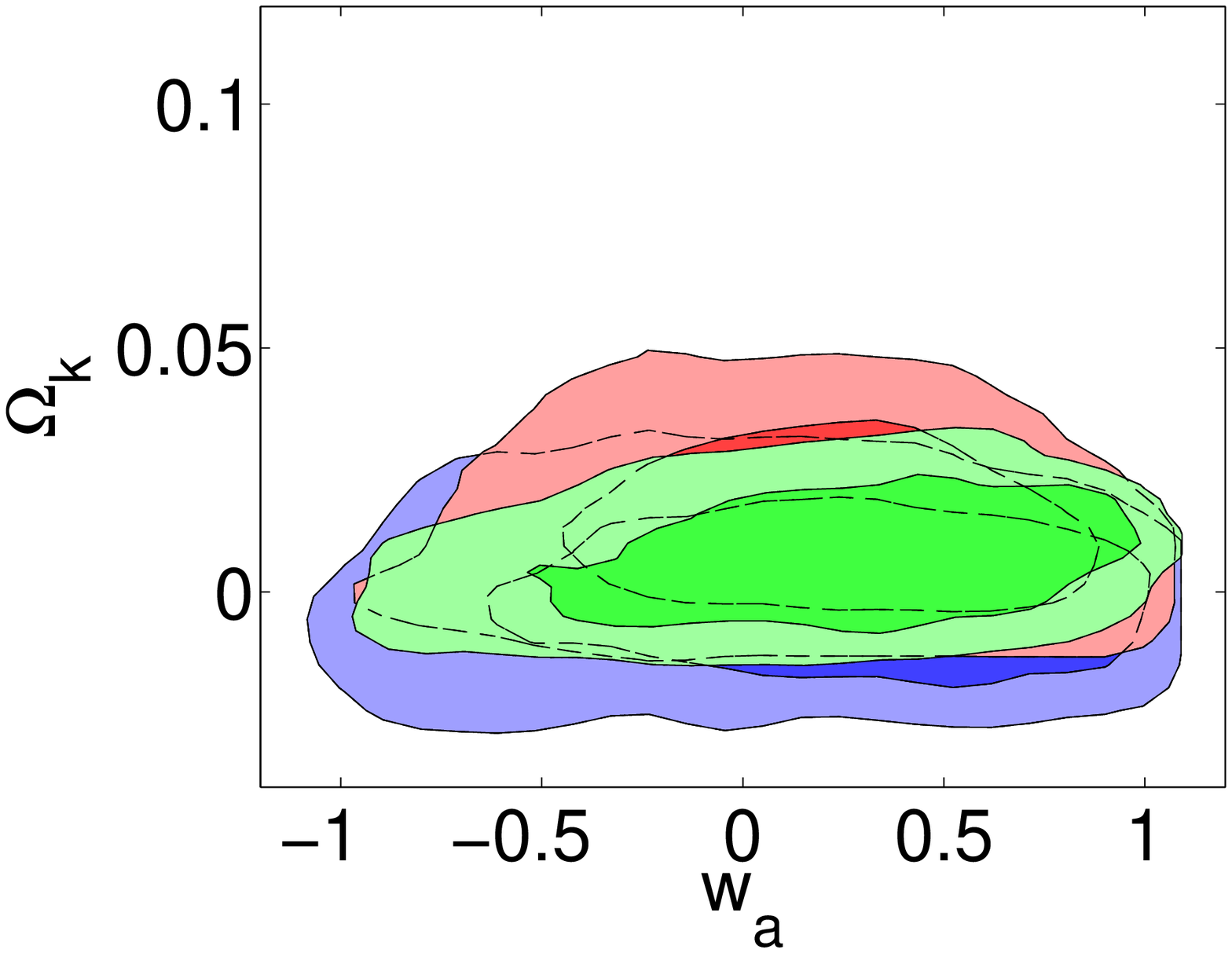} &
		 \includegraphics[width=7.5cm]{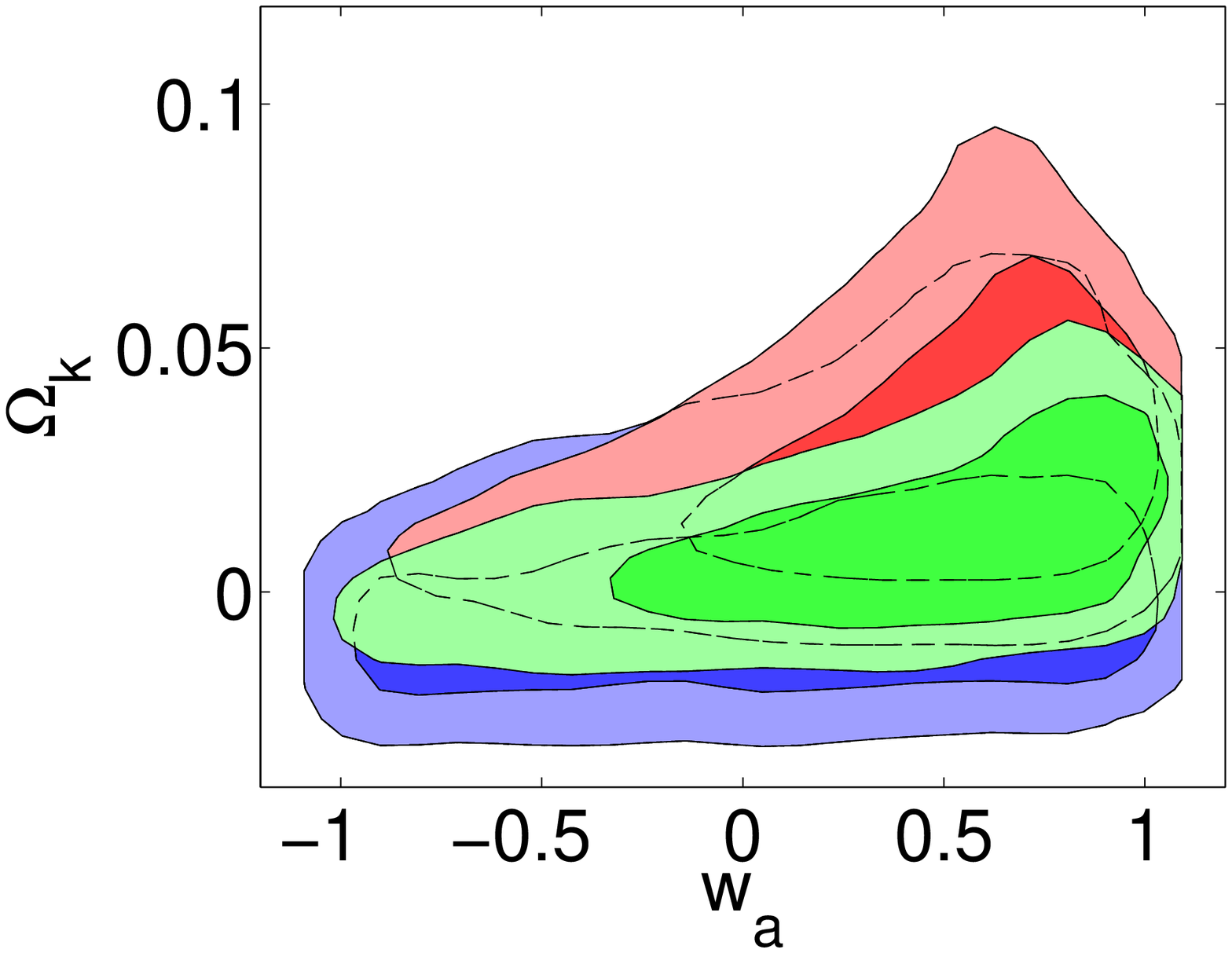} 
\end{tabular}
\caption{\it  $1$ and $2-\sigma$ constraints on the $w_{0}$-$\Omega_k$ plane for the full 9-parameters model. The blue, green and red contours depict run0, runI (both combined with SDSS BAO data) and runII results, respectively. Left panel: including  dark energy perturbations Right panel:  without including dark energy perturbations in the analysis.}
\label{fig:fig0b}
\end{center}
\end{figure}
Figure~\ref{fig:fig0b} shows the $1$ and $2\sigma$ constraints in the
$w_a$-$\Omega_k$ plane for the full 9-parameters model. The right
panel allows for perturbations in the dark energy while in the left
panel perturbations are switched off. 
 With current data we are unable to produce meaningful constraints on
 the $w_a$ parameter. Notice that the contours are closed by the prior
 imposed ($-1<w_a<1$).  It is also evident that the effect of the dark
 energy perturbations is important if the universe is open,
 i.e. $\Omega_k>0$. Analogously to what happens  in the flat, constant $w$
   case~\footnote{The authors of Ref.~\cite{SpergelWMAP06} conclude
     that, when the dark energy perturbations are considered, the (tiny)
     extra dark energy clustering reduces the size of the quadrupole
     and therefore its power to extract $w$.}, the addition of the dark energy perturbations into the analysis 
   changes the allowed parameter space, lessening the constraints in the
   $w_a$-$\Omega_k$ plane. 
   
   It is well known that crossing the phantom divide $w=-1$ can
  lead to divergences in the dark energy perturbation
  equations~\cite{huinst,caldwell}, for a   fixed dark energy sound speed $c^2_s$.  Also, dark energy perturbations may not be well physically well defined for models where  $w<-1$; for example modified gravity models or k-essence, can yield effectively $w<-1$. Some works in the literature switch off the perturbations when $w<-1$.  \cite{creminelliw} argues that  for physical solutions and to avoid instabilities and ghosts, the sound speed  should be zero-- or small and negative- for $w<-1$. Here we follow the treatment   presented in Ref.~\cite{lewishu}, which provides a method to cross 
  $w = -1$ and avoids instabilities. In addition, in the absence of a fully motivated and specified dark energy model,  we  simply show both  cases, with and without perturbations,  to quantify their effect.

 \section{The $w-\Omega_k$ degeneracy and future BAO surveys}
\label{sec:secii}
Acoustic oscillations in the photon-baryon plasma are imprinted in
the matter distribution. These Baryon Acoustic Oscillations (BAO)
have been detected in the spatial distribution of galaxies by the
SDSS~\cite{sdssbao} and the 2dF Galaxy Reshift Survey~\cite{2dfbao,percival}. 
The oscillation pattern is characterized by a standard ruler, $s$, 
whose length is the distance sound can travel between the Big Bang 
and recombination and at which the correlation function of dark matter 
(and that of galaxies, clusters) should show a peak. 
Detecting this scale $s$ at different redshifts is  the major  goal of future galaxy surveys.

Therefore, the aim of a BAO survey is to measure the location of the baryonic peak  in the correlation function along ($s_\| = \Delta z$) and across ($s_\bot = \Delta \theta$) the line of sight. In the radial direction, the BAO directly measure the instantaneous expansion rate $H(z)$ through  $s=(c/H(z)) s_\|$, where $H(z)$ is given by the Friedmann equation,
\begin{equation}
 H^2(z)= H^2_{0}\left(\Omega_{m} (1+z)^3 + \Omega_k (1+z)^2 +\Omega_\Lambda exp \left(3 \int^z _{0}\frac{1+w(z^\prime)}{1+z^\prime} dz^\prime \right) \right)
\label{eq:hubble}
 \end{equation}
 where $\Omega_m$, $\Omega_{k}$ and $\Omega_\Lambda= 1-\Omega_{m} -\Omega_{k}$ are the energy density of the universe in the form of dark matter, spatial curvature and dark energy respectively.

 At each redshift, the measured angular (transverse) size of oscillations, $s_\bot$, corresponds to the physical size of the sound horizon, $s(z)=d_A (z)s_\bot$, where the angular diameter distance $d_A$ reads
 \begin{equation}
 d_A(z)=\frac{1}{H_0\  \sqrt{-\Omega_{k}}} \sin \left(\sqrt{-\Omega_{k}} \int^{z}_{0} dz^\prime \frac{H_0}{H(z^\prime)} \right)~,
 \end{equation}
 which is formally valid for all curvatures, and $H(z)$ is given by Eq.(\ref{eq:hubble}). 
The fact that, given sufficient redshift precision, a BAO survey can measure both the  $H(z)$ component ($s=(c/H(z)) s_\|$) and the transversal component offers a powerful consistency check:  the recovered $H(z)$ must agree with the recovered $d_A(z)$, which is an integral  of $1/H(z)$. This feature is the key to disentangle curvature from dark energy properties as we will illustrate next.

There are future large scale surveys such as BOSS\footnote{http://www.sdss3.org/cosmology.php}, Euclid\footnote{http://sci.esa.int/science-e/www/area/index.cfm?fareaid=102}, JDEM\footnote{http://jdem.gsfc.nasa.gov/} and LSST\footnote{http://www.lsst.org/} planned, which will
cover $\mathcal{O}(10000)$ square degrees of the sky and are expected
to extract the angular extent of the BAO signature and, redshift
precision allowing, also the  BAO
feature in the radial direction. Therefore, these future surveys are
expected to provide measurements of $d_A(z)$ (and, in many cases of
$H(z)$) in the $z\ltap 3 $ redshift interval. 
In the next subsection we discuss how  in principle, reconstructing in a model-independent way a time
dependent dark energy equation of state $w(z)$ from measurements   
of $H(z)$ and $d_A(z)$ independently, could help to identify the
presence of cosmic curvature~\cite{prev3}. However, when realistic errors
from future surveys are considered, this procedure fails.  
The expected errors in $d_{A}(z)$ and $H(z)$ can be estimated using,
for instance, the Fisher matrix approach presented in Ref.~\cite{sande}.
Similar results could be obtained using the errors forecasted in Ref.~\cite{blake}.

\subsection{Reconstructing $w(z)$ via measurements of $H(z)$ and
  $d_A(z)$}
We start by following Ref. \cite{prev3}.   Let us assume that the universe is such that there is a small curvature component and a cosmological constant. In such a universe the Hubble parameter is given by the expression:
\begin{equation}
H^2(z)= H^2_{0}\left(\Omega_{m} (1+z)^3 + \Omega_k (1+z)^2
  +\Omega_\Lambda\right)~.
\label{eq:hubble2}
\end{equation}

\begin{figure}[!ht]
\begin{center}
 \begin{tabular}{c c} 
  \includegraphics[width=7.5cm]{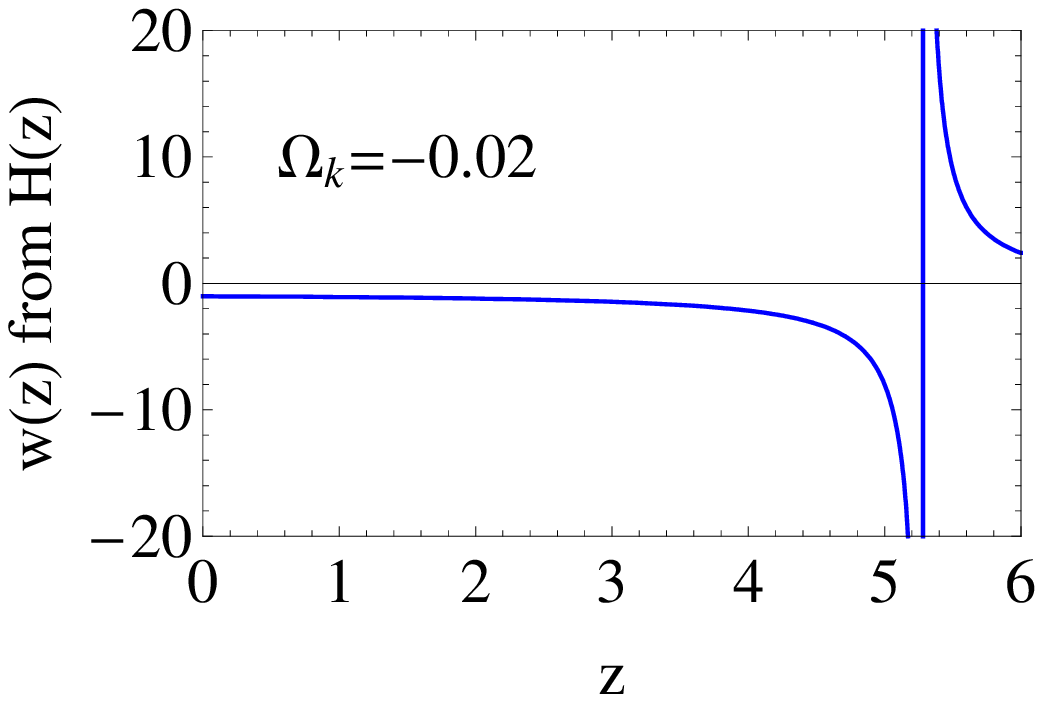}&
  \includegraphics[width=7.5cm]{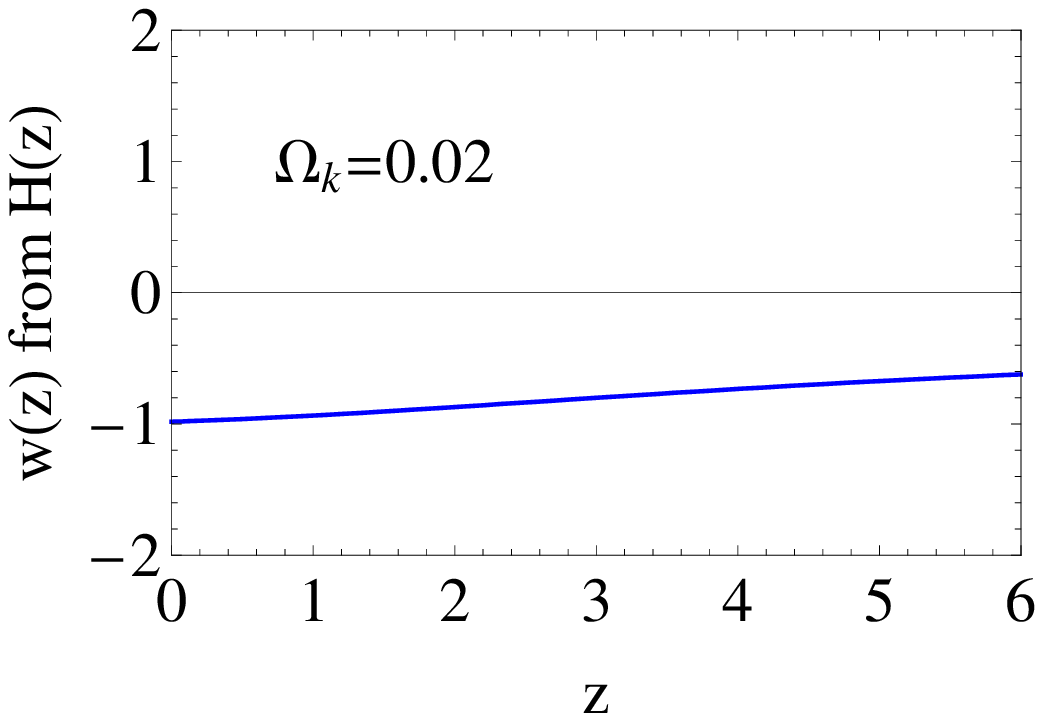}\\
  \includegraphics[width=7.5cm]{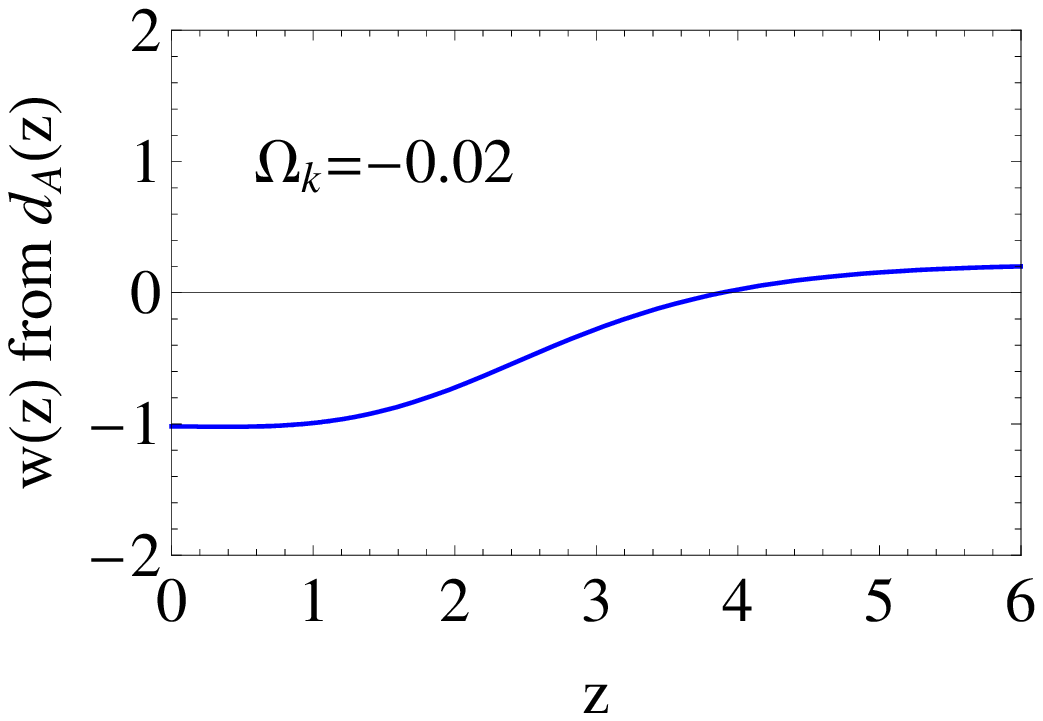}&
  \includegraphics[width=7.5cm]{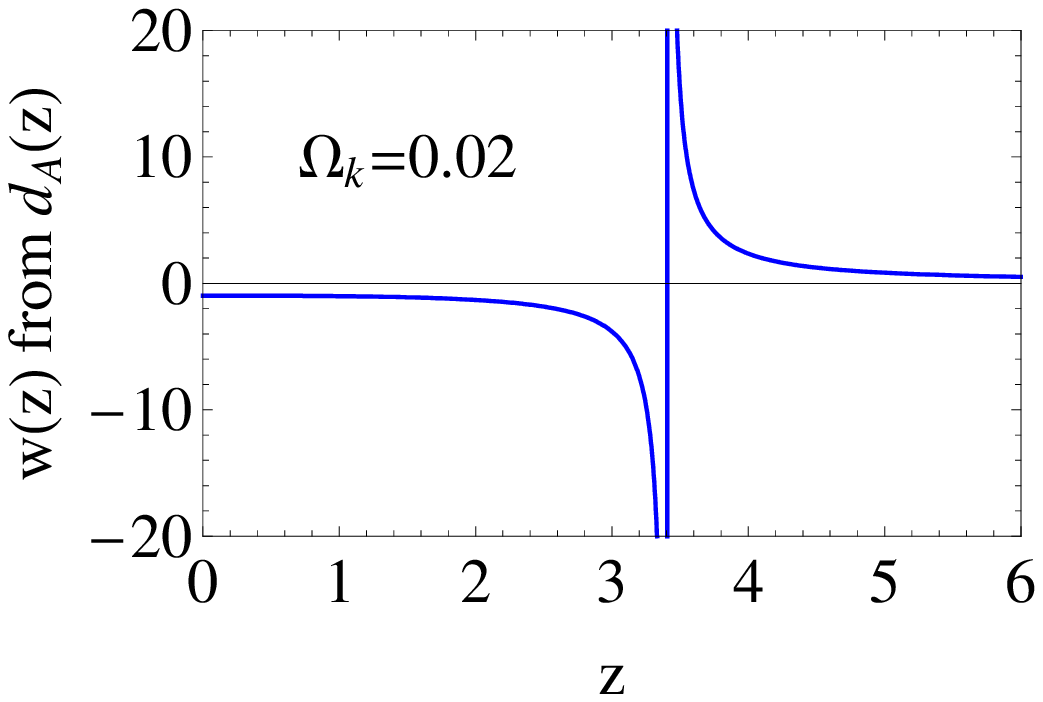}\\
 \end{tabular}
\caption{\it In an ideal case with perfect   measurements for the radial and angular BAO location as function of redshift, the inferred $w(z)$ from $H(z)$ and $d_A(z)$ under the assumption of flatness do not coincide in the presence of curvature. The top panels show the $w(z)$ inferred from $H(z)$: on the left  is the case where a LCDM negatively curved universe is erroneously assumed to be flat and on the right is the negatively curved  case.
The bottom panels show the equivalent situation for when $w(z)$ is inferred from $d_A(z)$. In all panels $|\Omega_k|=0.02$.}

\label{fig:wtheo}
\end{center} 
\end{figure}

If $H(z)$ could be perfectly measured, and used to  reconstruct $w(z)$ but assuming zero curvature, the inferred $w(z)$ would be the following function  of the Hubble parameter and its first derivative ~\cite{prev3}

\begin{equation}
w_H(z)=-\frac{1}{3}\frac{2(1+z)HH'-3H^2}{H_0^2\Omega_m (1+z)^3-H^2}~,
\label{eq:wH}
\end{equation}

where $'=\mathrm{d}/\mathrm{d}z$ and the Hubble parameter $H$ is given by Eq.(\ref{eq:hubble2}). 

If one focused instead on the angular diameter distance and its first and second derivatives, the $w(z)$ reconstructed would be given by 

\begin{equation}
w_{d_{A}}(z)= \frac{-3 d_{A}' - 2 (1+z) d_{A}''}{3\left( 1 - (1+z)^3 \Omega_{ m}d_{A}^{'2} \right) d_{A}}
\label{eq:wda}
\end{equation}

Figure \ref{fig:wtheo} depicts the $w(z)$ that would be inferred after
substituting in Eq.(\ref{eq:wH}) and in Eq.(\ref{eq:wda}) the value of
the Hubble parameter, the value of the angular diameter distance and
their derivatives versus redshift for a non flat universe with a cosmological
constant. 

For the example illustrated in Figs.~\ref{fig:wtheo}, we assume
$|\Omega_k|=0.02$.
Notice that, for negative curvature, while the $w(z)$ inferred from $H(z)$ and
$H'(z)$ tends to values $w<-1$ as the redshift increases, the $w(z)$
reconstructed from the angular diameter distance and its derivatives
tends to values $w>-1$. For positive curvature the behaviour of the
$w(z)$ reconstructed from the angular diameter distance and the $w(z)$
inferred from $H(z)$ is the opposite.

An incorrect assumption about the Universe geometry would therefore show up as an inconsistency between the radial and transverse BAO. After combining measurements of $H(z)$ and $d_A(z)$ one would naively expect to break the $w(z)$-curvature degeneracy and to reconstruct an equation of state which resembles the underlying 
\emph{true} cosmology, i.e. $w=-1$.

Perhaps the most attractive feature to attempt such a reconstruction
lies in the fact that the inferred $w_H(z)$ and $w_{d_{A}}(z) $ exhibit a
resonant-like behaviour, i.e. only one of the reconstructed $w$'s will
diverge depending on the sign of the curvature and the position of the
pole will signal the size of this curvature.
A negative curvature will be indicated by  a resonant behaviour in the
inferred $w_H(z)$ with a pole at a redshift of
\begin{equation}
z= \sqrt{- \Omega_{\Lambda}/\Omega_{ k}} - 1~,
\end{equation}
while a positive curvature will make  $w_{d_{A}} \rightarrow \infty$ at $z$ satisfying the condition
\begin{equation}
H^2_0(1+z)^3 \Omega_{m} \cosh^{2}\left( \sqrt{\Omega_{k}} \int \frac{H_{0}}{H(z')} dz'\right) = H^2(z)\,.
\end{equation}

Therefore, one can be lead to believe that such a striking feature cannot be
missed and envision that the degeneracy between $w(z)$ and $\Omega_{k}$  can easily be lifted. Unfortunately, for realistically achievable constraints on $H(z)$ and $d_A(z)$ this will not  be the case as we will show next.

\begin{figure}[!ht]
\begin{center}
 \begin{tabular}{c c} 
  \includegraphics[width=7.5cm]{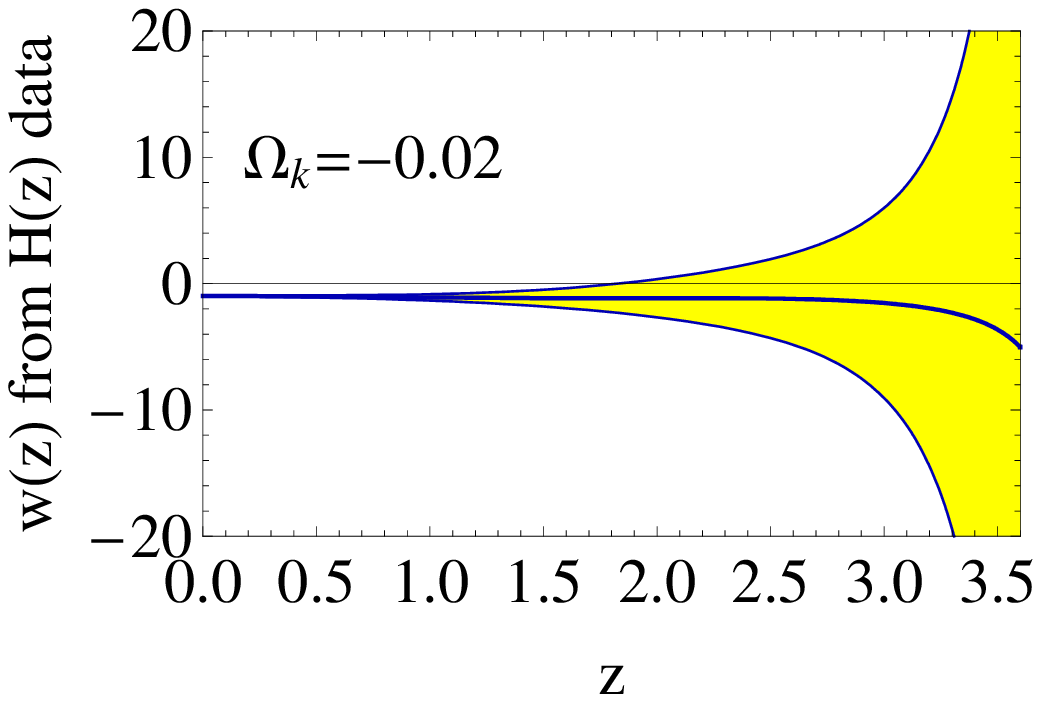}&
  \includegraphics[width=7.5cm]{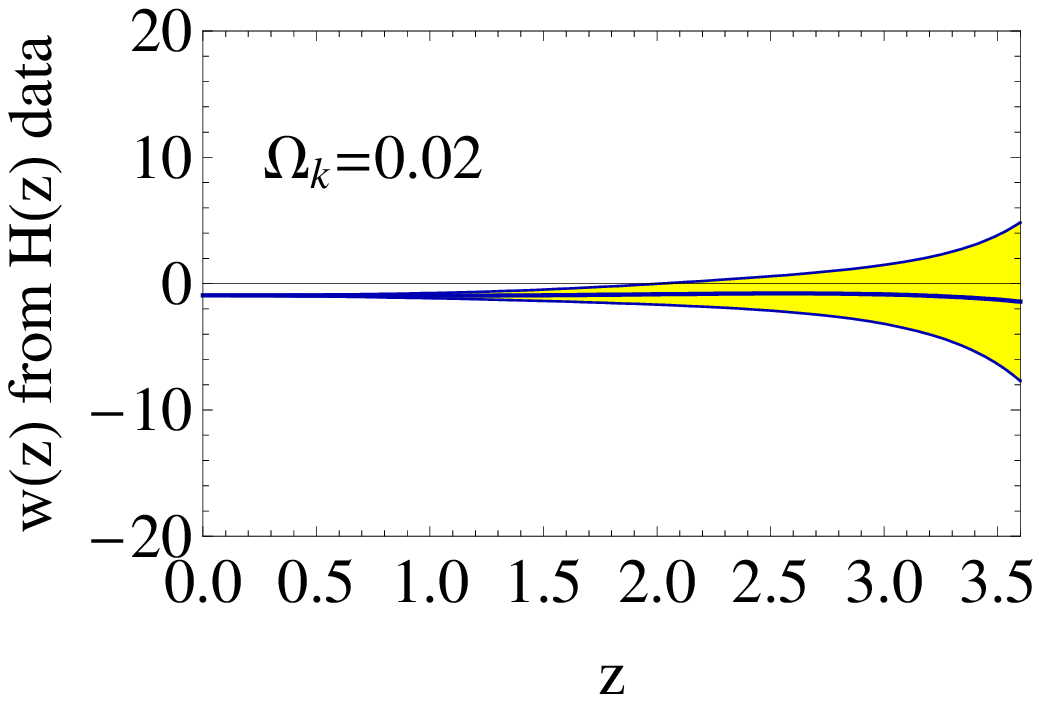}\\
  \includegraphics[width=7.5cm]{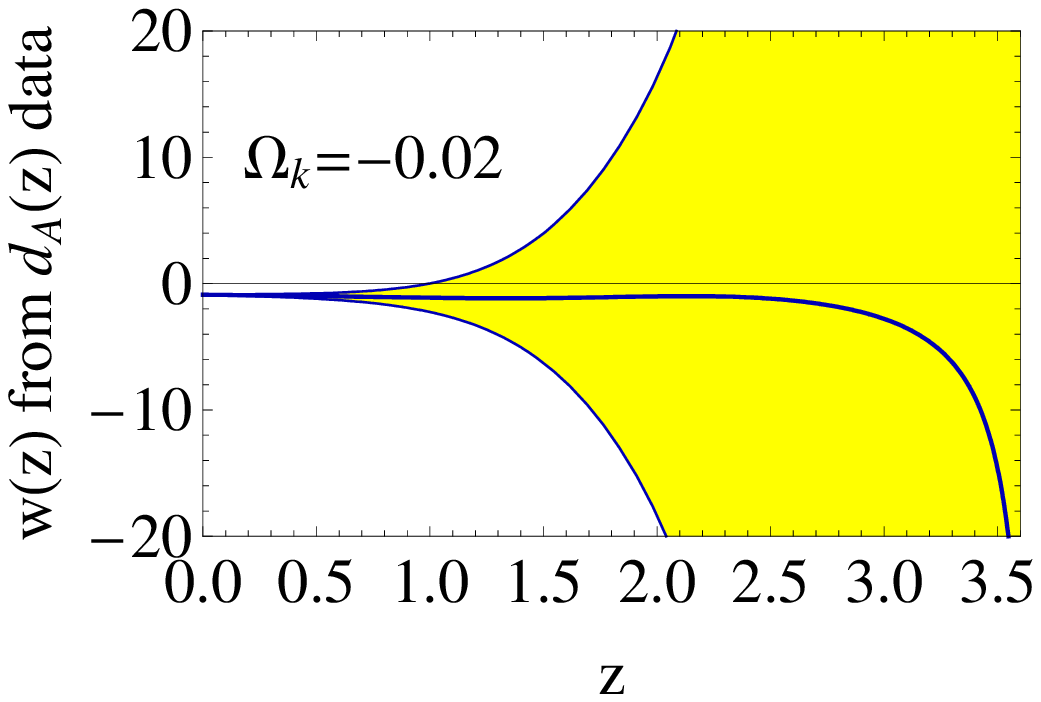}&
  \includegraphics[width=7.5cm]{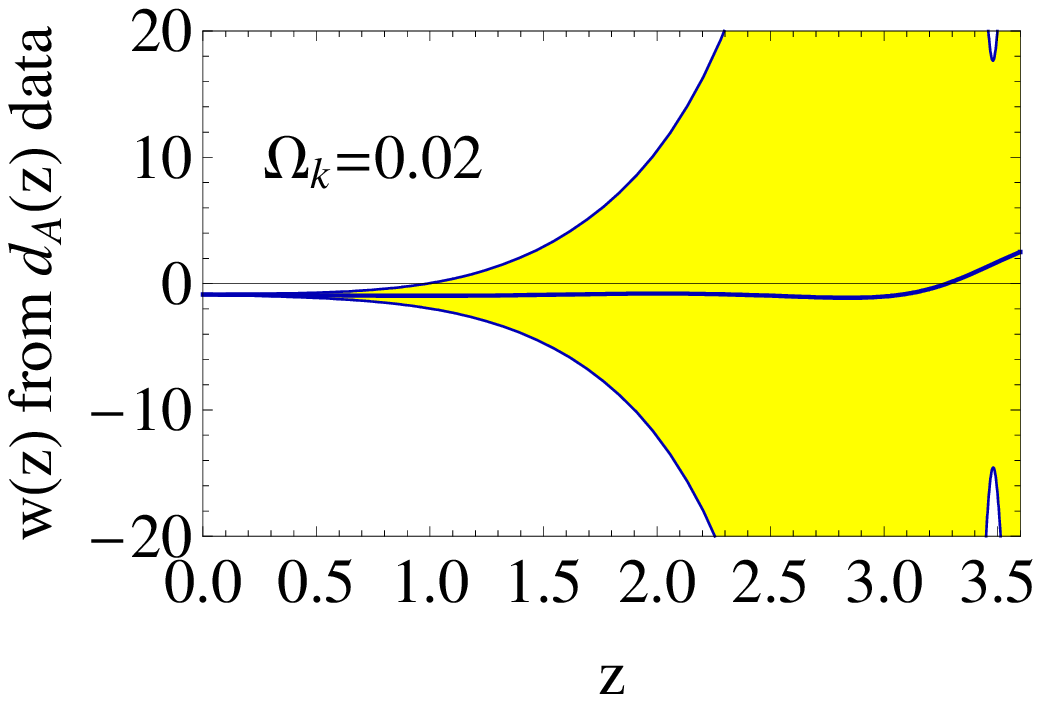}\\
 \end{tabular}
\caption[]{\textit{ The top (bottom) panels depict the reconstructed $w(z)$ from 
 $H(z)$ ($d_A(z)$)  best fit polynomial, see 
Eqs.~(\ref{eq:wH}) and (\ref{eq:wda}). 
The left (right) figures illustrate the case of a
negative (positive) curvature $ \mid \Omega_k \mid=0.02$.}}
\label{fig:wrec}
\end{center} 
\end{figure}

In order to mimic future $H(z)$ and $d_A(z)$ data, we have assumed a very 
optimistic survey, similar to an LSST-type survey but with no photo-z errors, 
with a volume of $30000$ squared degrees and a
redshift range from $z=0.3$ up to $z=3.6$, in bins of $\Delta z=0.1$
width. The mean galaxy density is chosen to be $n=3\times 10^{-3}$. 
After generating  mock data for $H(z)$, $d_{A}(z)$ and their errors
(computed with the Seo and Eisenstein procedure, see ~\cite{sande}), we fit the
mock data to a 3rd (4th) grade polynomial for $H(z)$ $(d_{A}(z))$. This choice is motivated as follows. Even with extremely high-quality future data, only a reduced number of dark energy parameters can ever be measured e.g.,\cite{LinderHuterer}. A general and flexible fitting function is thus a polynomial. Eigenmodes or bin-based approaches \cite{bins,principalcomponents} would effectively impose a drastic smoothing, erasing the signal even more. 
Here, the role of this parameterization is primarily to quantify the size of the error-bars of $w_H$ and $w_{d_A}$ achievable from future surveys. As it is clear  
from Fig.~\ref{fig:wrec} the error-bars are much larger than the``signal'' making the  
conclusions of this section rather insensitive to the type of parameterization used for $w(z)$.

With these polynomials we  reconstruct $w_H(z)$ and 
$w_{d_{A}}(z)$ together with their errors and we present the results
in Fig.~\ref{fig:wrec}. For $w_{m}=\Omega_{m} h^2$ we assume a $2\%$
error, as expected from Planck data~\cite{planck}. 
If the curvature is negative, 
the reconstructed $w(z)$ from $H(z)$ should have a divergence, while the
$w(z)$ reconstructed from $d_{A}(z)$ should not diverge.
The different behaviour for these two $w(z)$'s was already pointed
out by the authors of Ref.~\cite{prev3}. 

However, in practice, this reconstruction procedure is
not good enough to reproduce the divergence that would signal the presence of curvature.
Indeed, as can be seen from Fig.~\ref{fig:wrec}, the errors with which the coefficients of the fitting polynomials would be reconstructed,
even with the most optimistic survey assumed here, yield an allowed $w(z)$ region too large to see the expected signal.
Moreover, the order of the fitting polynomials assumed is too low
to track precisely enough the behaviour of $H(z)$ and $d_A(z)$ to the high redshifts at which the divergence occurs. Indeed, we find that
for $z>3$ the fit is not reliable, explaining why the best fit curves in Fig.~\ref{fig:wrec} may miss (or find) a divergence when 
their theoretical counterparts of Fig.~\ref{fig:wtheo} (do not) display it.

We therefore conclude that, at least  with the polynomial reconstruction method followed here, curvature and dynamical dark energy can not be disentangled by using the functions $w_H(z)$ and $w_{d_a}(z)$. It could be interesting to see if more refined parametric reconstruction approaches, or even non-parametric methods (see eg. \cite{Sahni:2006pa}), could alleviate the $\Omega_k-w(z)$ degeneracy. This, at least at first sight, seems unlikely for forthcoming experiments: the error-bars on the reconstruction seems to be much larger than the signal in the redshift range accessible to future galaxy surveys. We can however still hope to be able to use a simple parameterization of $w(z)$ and to separate curvature from dark energy via a likelihood analysis from  future BAO data.
 
\section{Future constraints}
\label{sec:seciii}
We discuss here the forecasted errors on the dark energy equation of
state expected from future surveys, in particular a survey with the characteristics of Planck \footnote{http://www.esa.int/Planck} for the CMB 
and surveys with characteristics not too dissimilar from  those of  BOSS, Euclid/JDEM and LSST for BAO surveys, allowing for non-zero spatial curvature (for a related work see \cite{McDonald:2006qs}).

The parameterization chosen for $w(z)$ is given
by Eq.~(\ref{eq:eosp2}). We will assume that the fiducial model is $\Lambda$CDM, that is, $w_0 = -1$ and $w_a=0$, but
we will leave $w_{0}$ and $w_{a}$ as free parameters in the fit. 

The forecasted errors are estimated using the Fisher matrix
formalism. For the CMB data, we have computed the Fisher matrix for  a full-sky CMB experiment with the noise and resolution characteristics of the Planck survey. We used the following parameters for the Fisher matrix describing Planck data:
\begin{equation}
  \theta=\left\{w_{b}, w_{dm}, \theta_{CMB},  \Omega_k, \tau, n_s, \alpha, A_s \right\}~,
\end{equation}
where $w_{b}=\Omega_{b} h^2$ and $w_{dm}=\Omega_{dm} h^2$ are the
physical baryon and dark matter densities respectively, $\theta_{CMB}$ is proportional to the ratio of the sound horizon to
the angular diameter distance, $\tau$ is the reionisation optical
depth, $n_s$ is the scalar spectral index, $\alpha$ is the running of
the scalar spectral index and $A_s$ the scalar amplitude.

We combine the CMB Fisher matrix with the BAO one for the large scale structure surveys.
We build the Fisher matrix assuming measurements of $H(z)s$ and
$d_{A}(z)/s$ in redshift bins of $0.1$ width 
where $s$ is the BAO scale, see Ref.~\cite{sande} for a 
detailed description of the method used to estimate the errors on 
$H(z)s$ and $d_{A}(z)/s$. 
The characteristics and redshift intervals of the different surveys 
are shown in Tab.~\ref{tab:tab1}. BOSS and Euclid/JDEM are
spectroscopic surveys and therefore the corresponding photo-z errors are set to zero.
We illustrate two possible photo-z errors for the LSST-type survey ($2\%$ and $5\%$), encompassing optimistic and more realistic expectations.

\begin{figure}[!ht]
  \begin{center}
 \begin{tabular}{c c} 
  \includegraphics[width=7.5cm]{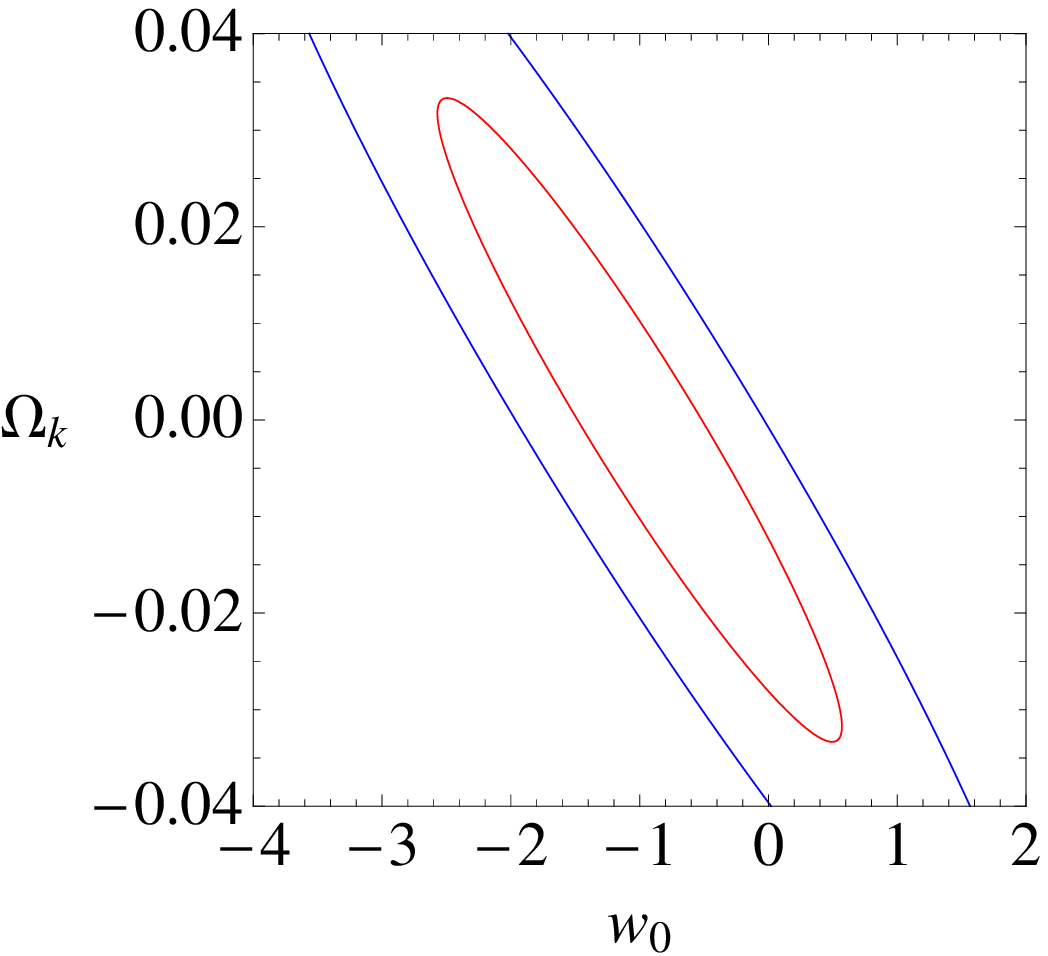}&
  \includegraphics[width=7.5cm]{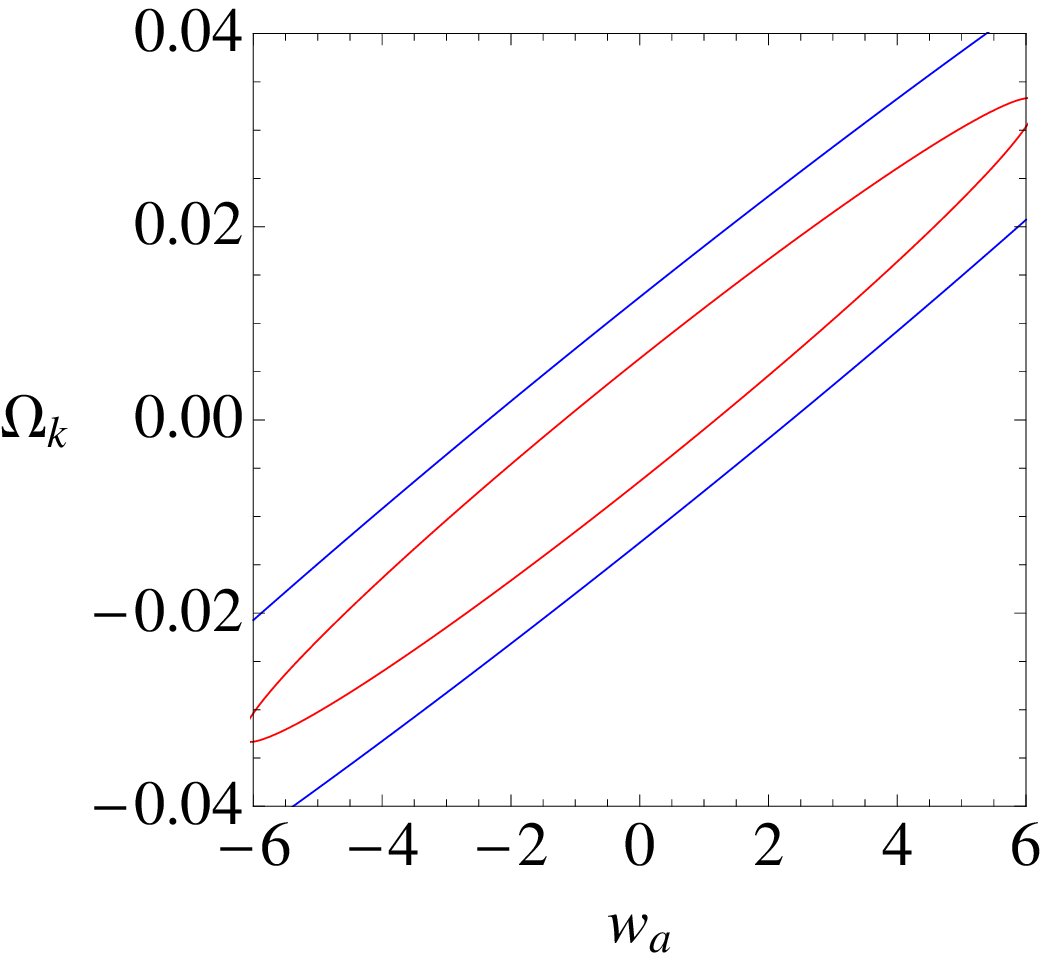}
 \end{tabular}
\caption{\textit{Left (right) panel: $1$ and $2-\sigma$ forecasted errors from
    the BOSS-type + Planck-type surveys  in the $w_{0}$-$\Omega_k$ ($w_a$-$\Omega_k$) plane.}}
\label{fig:wpboss}
\end{center} 
\end{figure}

\begin{figure}[!ht]
  \begin{center}
 \begin{tabular}{c c} 
  \includegraphics[width=7.5cm]{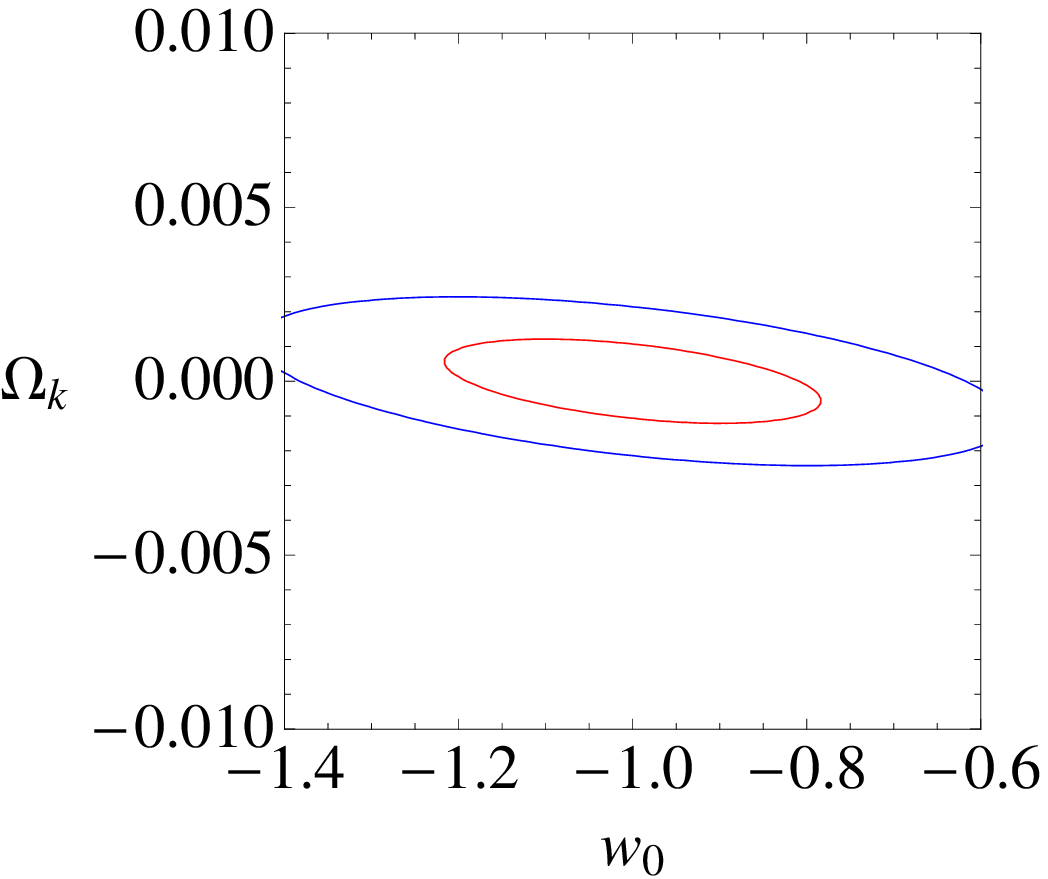}&
  \includegraphics[width=7.5cm]{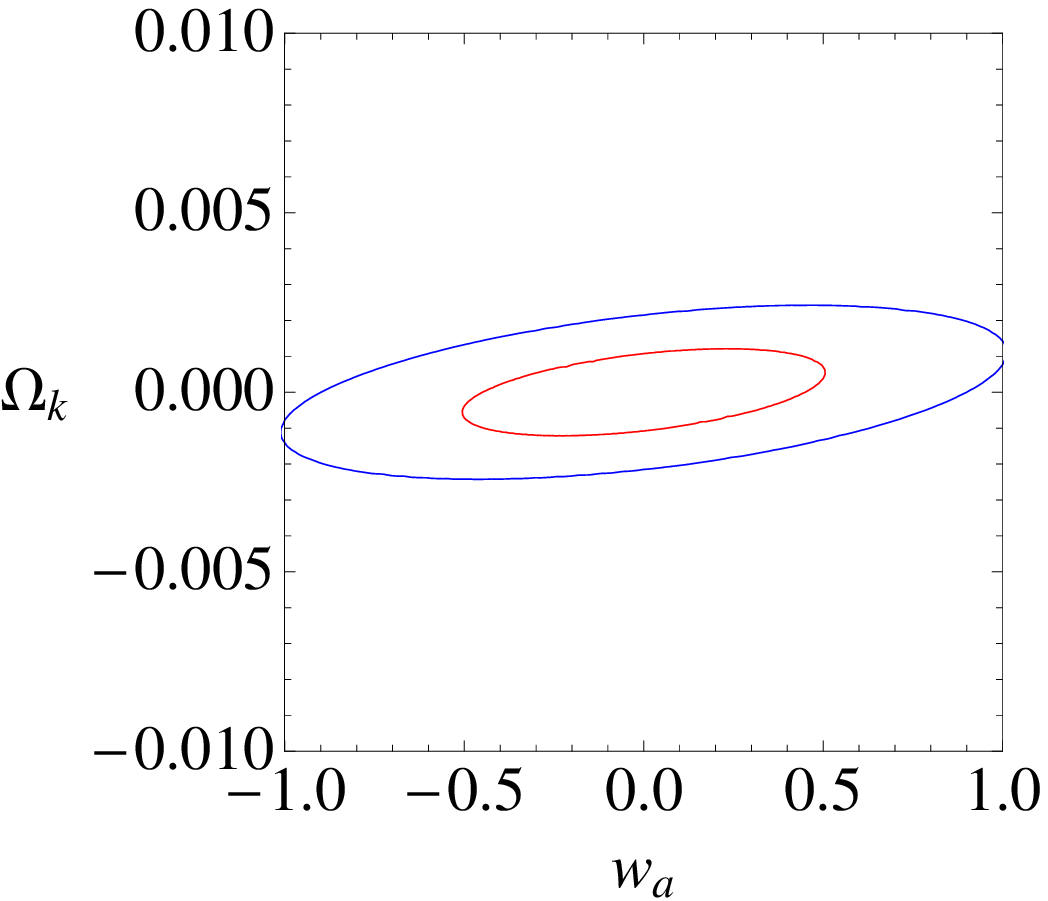}
 \end{tabular}
\caption{\textit{Left (right) panel: $1$ and $2-\sigma$ forecasted errors from
    the Euclid/JDEM-type + Planck-type surveys  on the $w_{0}$-$\Omega_k$ ($w_a$-$\Omega_k$) plane.}}
\label{fig:wpjdem}
\end{center} 
\end{figure}

\begin{figure}[!ht]
  \begin{center}
 \begin{tabular}{c c} 
  \includegraphics[width=7.5cm]{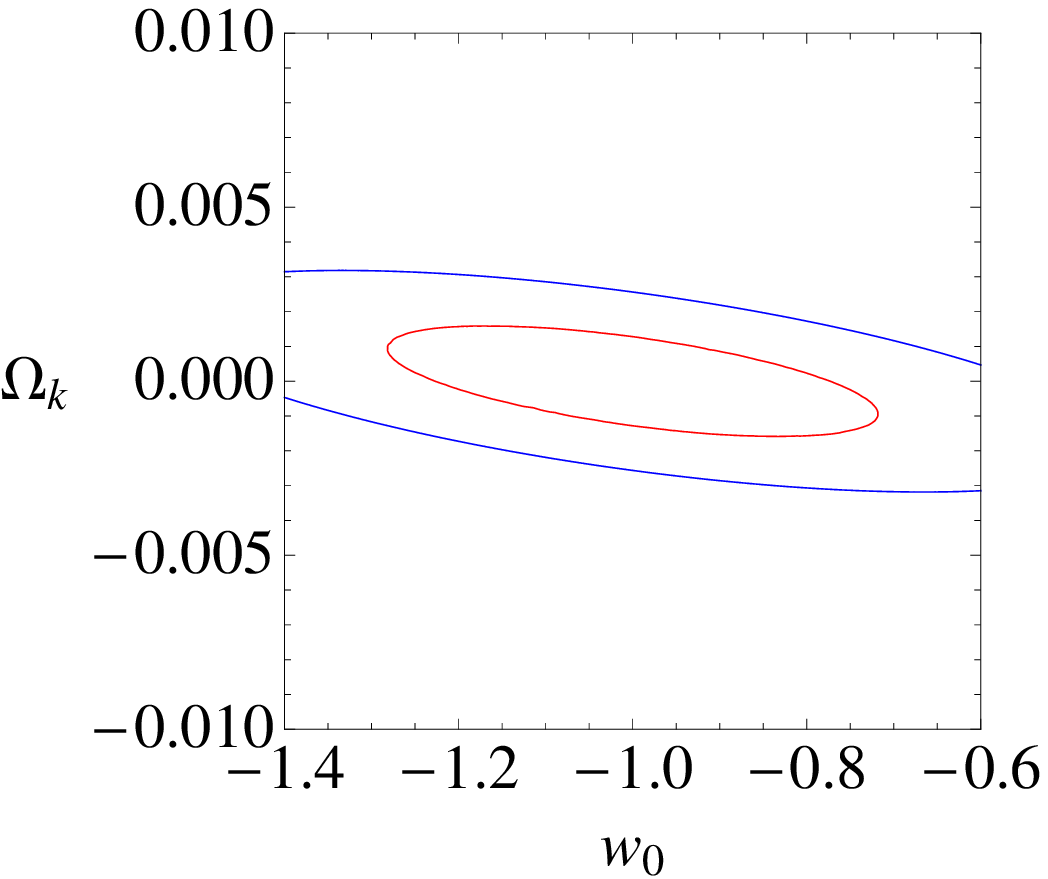}&
  \includegraphics[width=7.5cm]{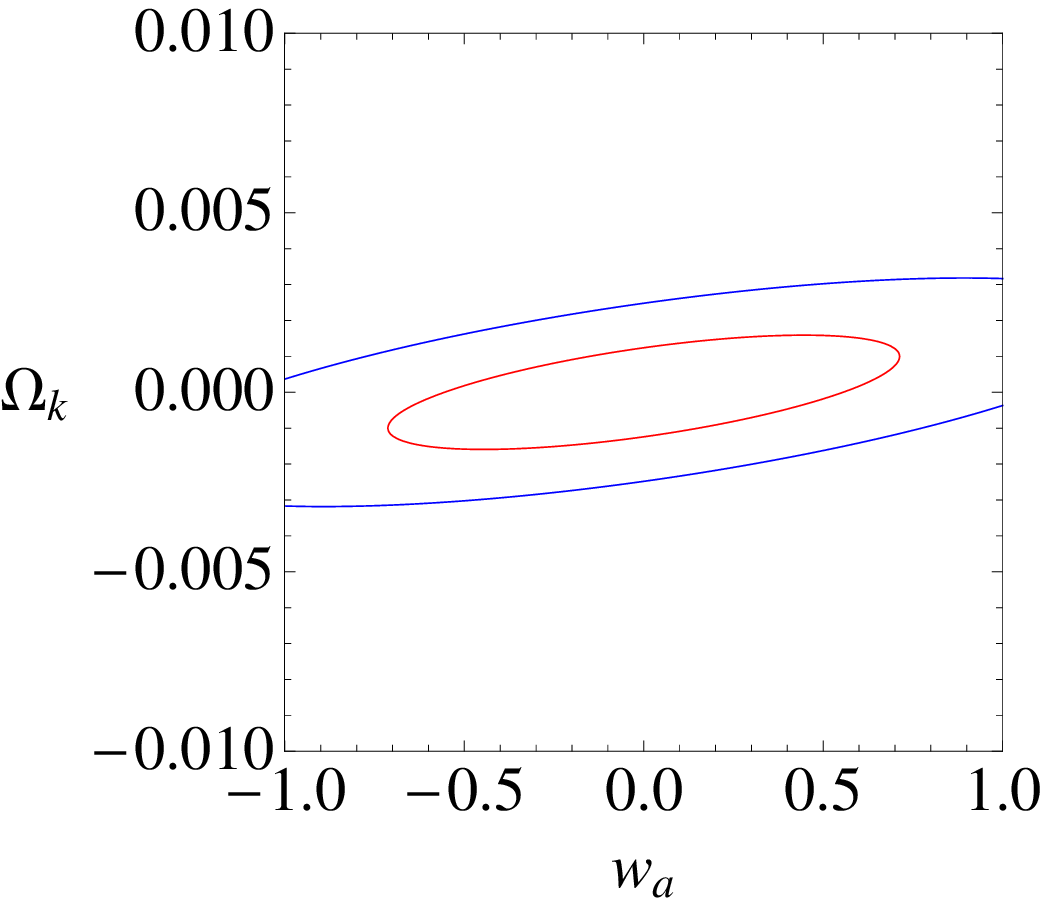}
 \end{tabular}
\caption{\textit{Left (right) panel: $1$ and $2-\sigma$ forecasted errors from
    the LSST + Planck surveys  on the $w_{0}$-$\Omega_k$
    ($w_a$-$\Omega_k$) plane. A $2\%$ photo-z error is assumed for the
  LSST survey.}}
\label{fig:wplsst2}
\end{center} 
\end{figure}

\begin{figure}[!ht]
  \begin{center}
 \begin{tabular}{c c} 
  \includegraphics[width=7.5cm]{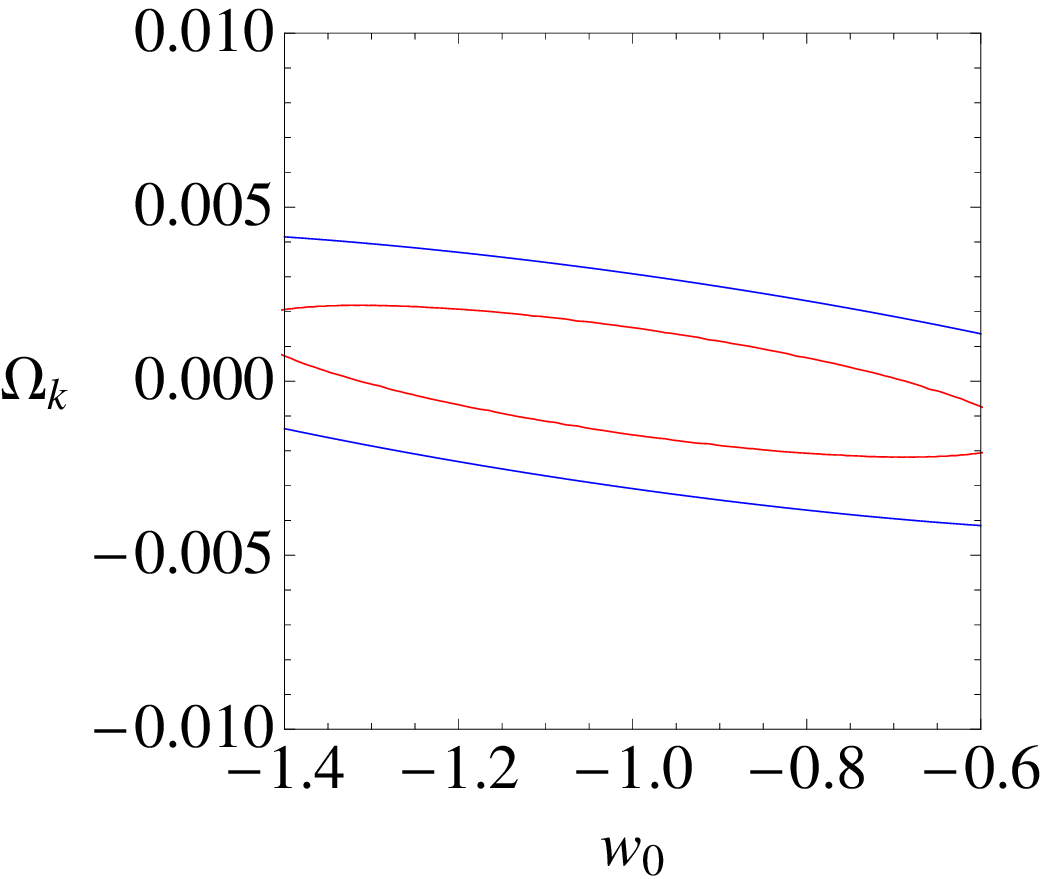}&
  \includegraphics[width=7.5cm]{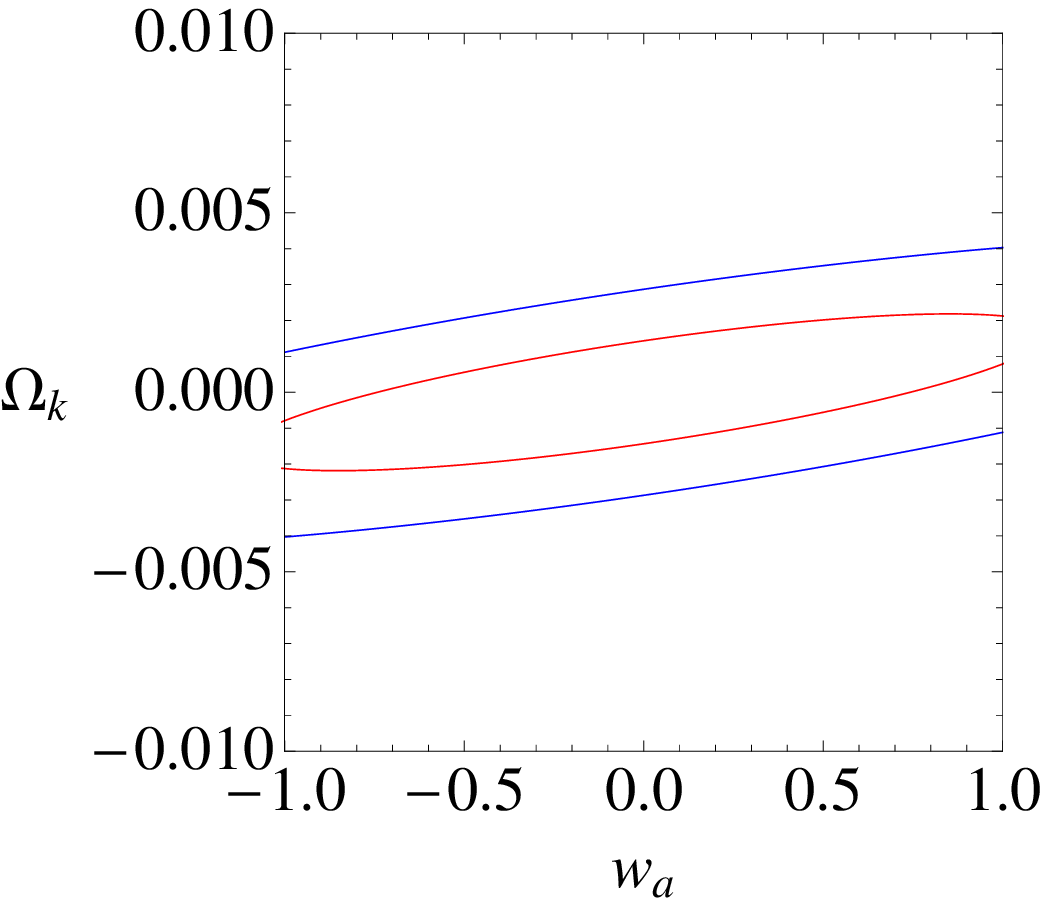}
 \end{tabular}
\caption{\textit{Left (right) panel: $1$ and $2-\sigma$ forecasted errors from
    the LSST + Planck surveys  on the $w_{0}$-$\Omega_k$
    ($w_a$-$\Omega_k$) plane. A $5\%$ photo-z error is assumed for the
  LSST survey.}}
\label{fig:wplsst5}
\end{center} 
\end{figure}

\begin{table}[bt!]
\begin{center}
\begin{tabular}{|c||c|c|c|c|} \hline
Survey & $n (h/Mpc)^3$ & Area (square degrees)& Redshift range &$\sigma_z$\\
\hline \hline
BOSS& $3\times10^{-4}$& $10000$& $0.1-0.7$ & $0$\\
 \hline
Euclid/JDEM& $1.9\times10^{-3}$& $20000$& $0.7-2.0$ & $0$\\
 \hline
LSST& $3\times10^{-3}$& $30000$& $0.3-3.6$ & $2\%$, $5\%$\\
\hline \hline
\end{tabular}
\caption{\label{tab:tab1} Mean galaxy density, covered area, redshift
  range and photo-z error of the different surveys considered here.} 
\end{center}
\end{table}
In order to produce forecasts for the dark energy parameters and the
spatial curvature we combine the two Fisher matrices after performing a transformation to the 
following set of parameters:$(H_0,\Omega_k, w_{b}, w_{dm}, w_0, w_a, \tau, n_s, \alpha, A_s)$. Notice that only
the CMB Fisher matrix contains information regarding the last four parameters 
$\tau$, $n_s$, $\alpha$, and $A_s$. Finally we extract the contours 
in the $w_{0}$-$\Omega_k$, $w_a$-$\Omega_k$ planes depicted in
Figs.~\ref{fig:wpboss},~\ref{fig:wpjdem},~\ref{fig:wplsst2} and \ref{fig:wplsst5}, marginalising over the other parameters. 


Surveys limited to low-redshifts will not have enough statistical power to lift the dark-energy-curvature  degeneracy,  see Figs.~\ref{fig:wpboss}.
The situation may improve if future  SNIa luminosity distance data  
were to be included. However SNe data effectively add statistical  
power only to the $d_A$ constraint not to the $H(z)$ one. In fact $d_L$ and  
$d_a$ are both integrals of $H(z)$ and differ only by a normalization  
factor of $(1+z)^2$. As shown by Ref.~\cite{FernandezMartinez:2008hw} $H(z)$ is the key observable in  
disentangling $w(z)$ from $\Omega_k$.

Euclid/JDEM-type data, with extended volume and redshift coverage, will improve significantly the
current curvature constraints, see Figs.~\ref{fig:wpjdem}.  
A similar improvement could be provided by the photometric LSST survey
if its photo-z $\sim 2\%$, see Figs.~\ref{fig:wplsst2}. The curvature
could be constrained to the $\sim 0.001$ level and the
constraints on the dark energy parameters could be improved by almost
a factor four. Moreover, with its extended redshift coverage, such a survey will yield  almost no residual correlation between the dark
energy equation of state parameters and curvature, as can be seen in Figs.~\ref{fig:wpjdem} and~\ref{fig:wplsst2}.
If the photo-z error is higher, $\sigma_z\sim 5\%$, the error 
in the curvature is still at the $\sim 0.001$ level but the error on the
dark energy parameters $w_0$ and $w_a$ increases considerably, 
see Figs.~\ref{fig:wplsst5}. The reason for that is because a higher 
photo-z error will suppress exponentially the radial BAO modes and 
therefore the information on $H(z)$ (which is crucial for the measurement of
$w(z)$) will be completely lost.

\begin{figure}[]
 \begin{tabular}{c c c} 
  \hspace{-1.0cm}
  \includegraphics[width=5.3cm]{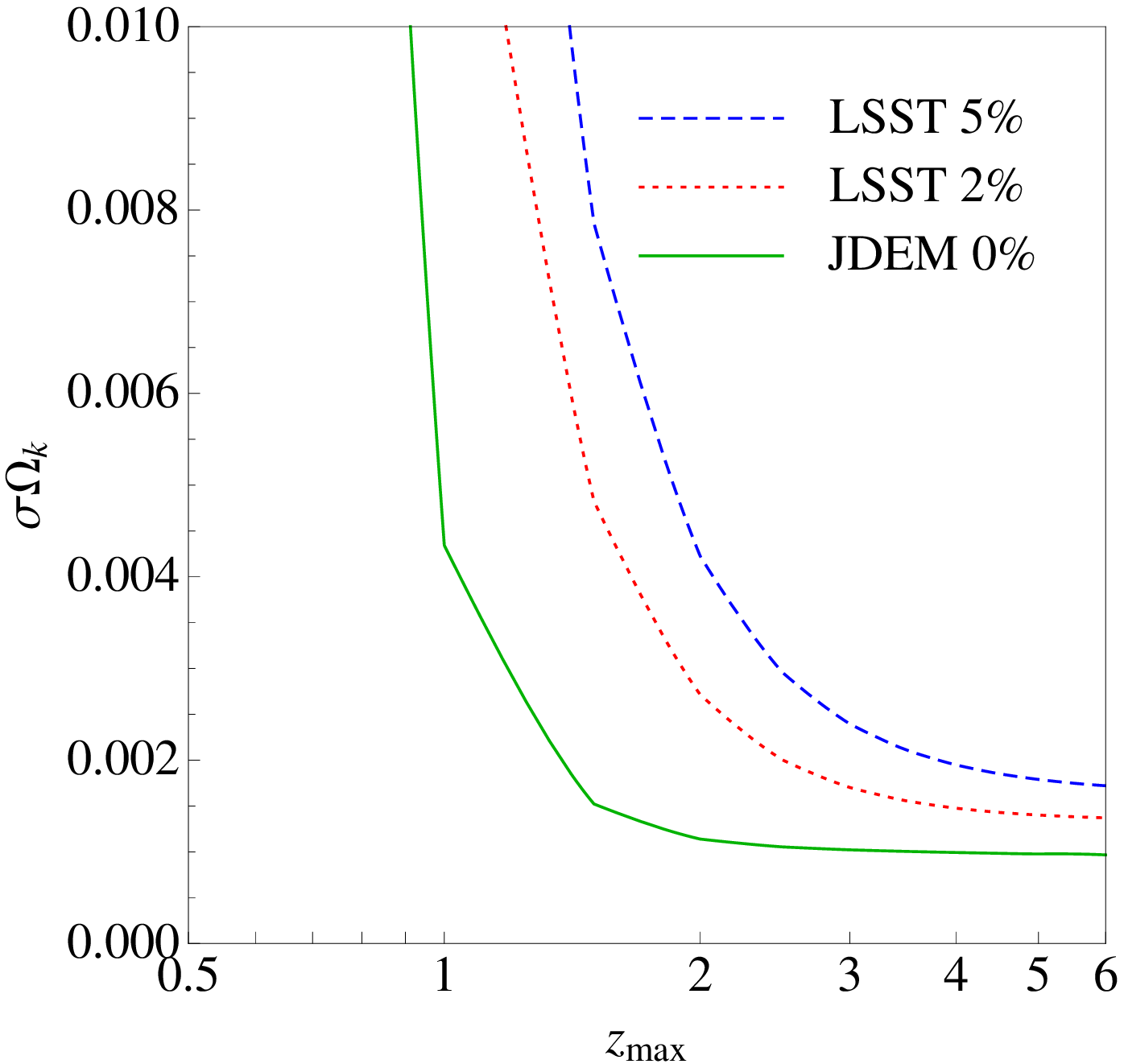}&
  \includegraphics[width=5.0cm]{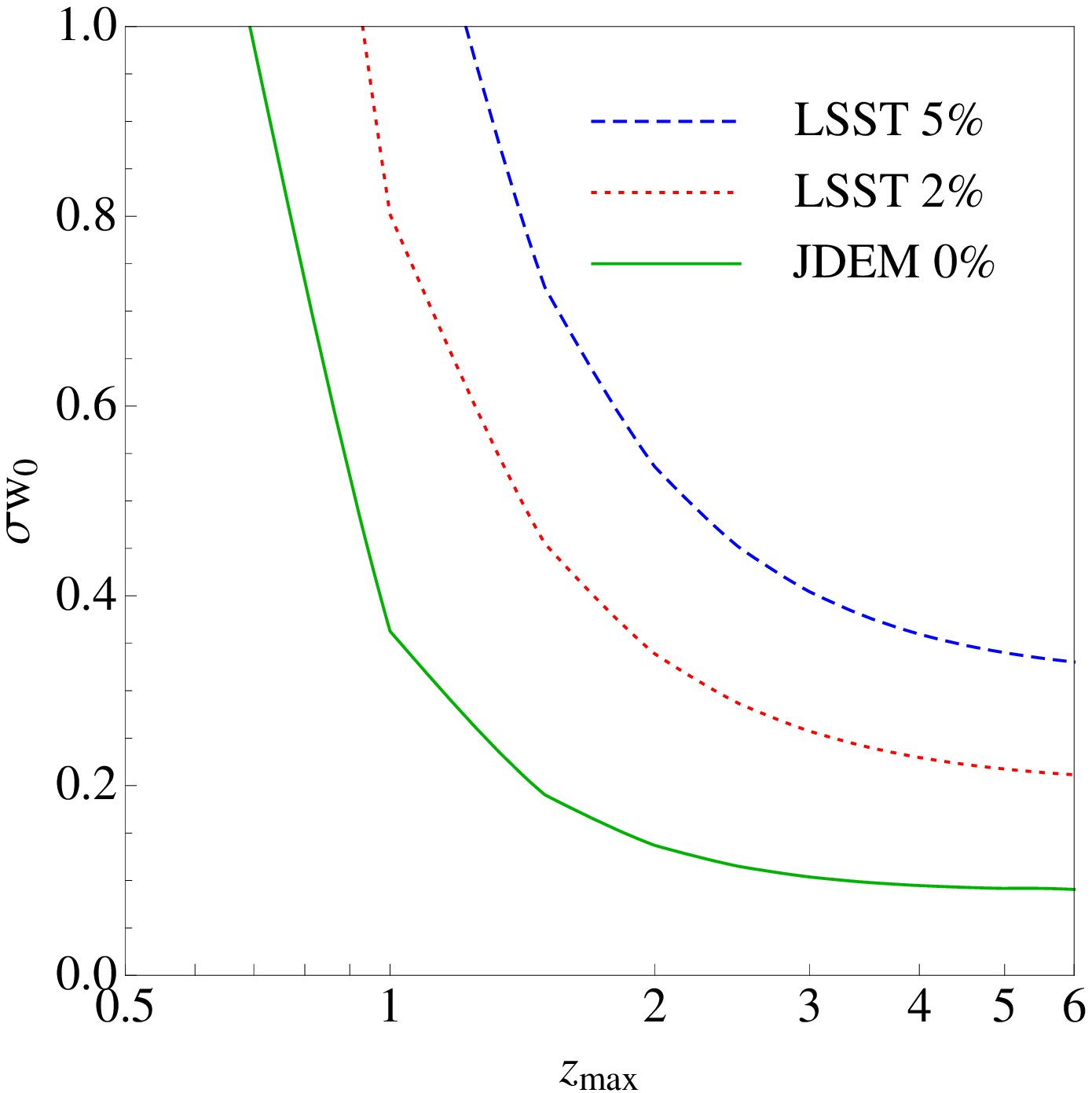}&
  \includegraphics[width=4.8cm]{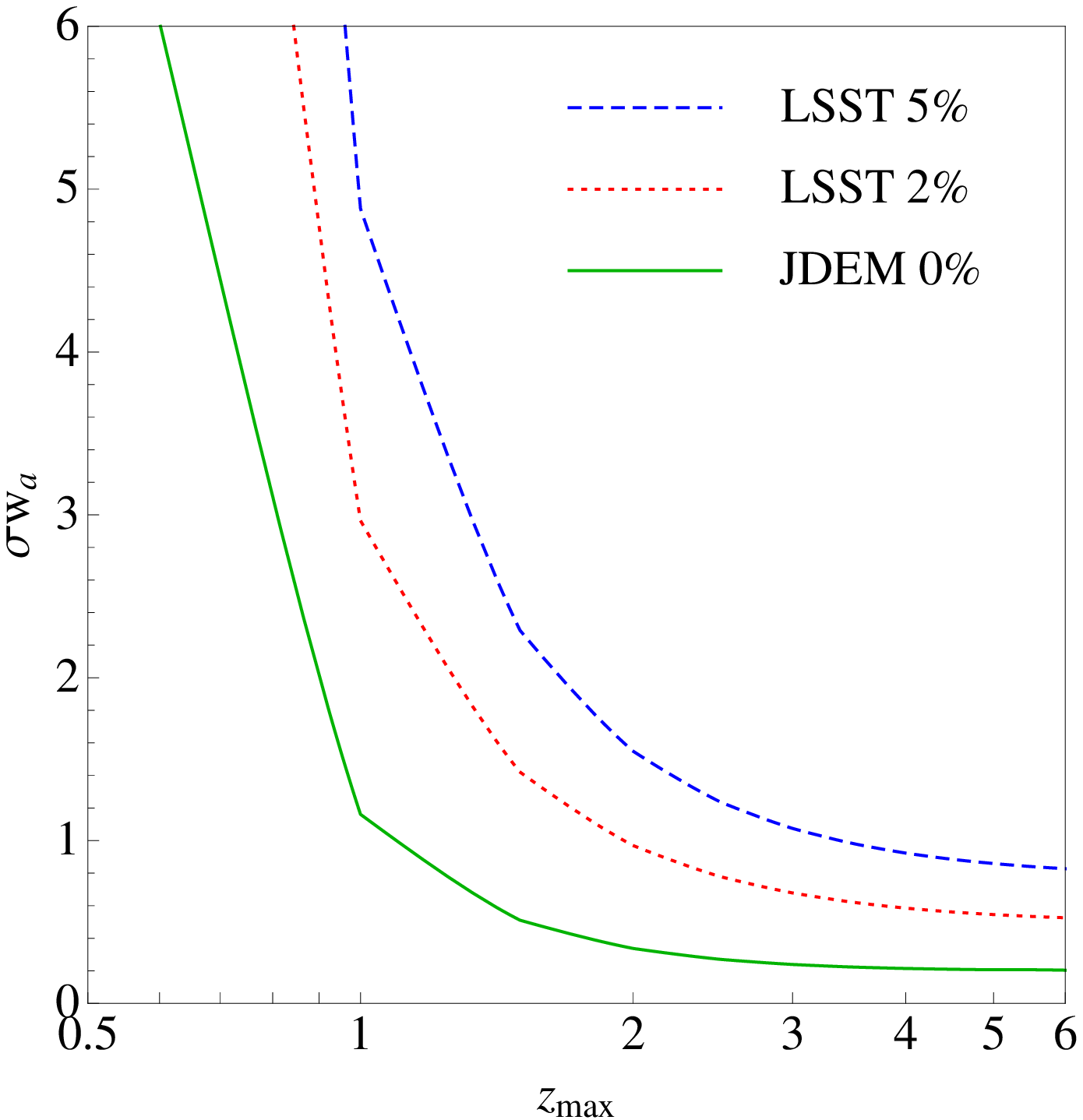}
 \end{tabular}
\caption{\textit{$1\sigma$ marginalised error on $w_0$, $w_a$
and $\Omega_{k}$ expected from Euclid/JDEM-type and LSST-type surveys versus the redshift coverage (maximum redshift).  For the
LSST-type survey, we illustrate the results assuming two different photo-z
errors, $2\%$ and $5\%$.}}
\label{fig:1sigma}
\end{figure}

In Figs.~\ref{fig:1sigma} we investigate the effect of a survey's redshift coverage on the parameter's errors. We show the $1\sigma$ marginalised error on $w_0$, $w_a$
and $\Omega_{k}$ expected from Euclid/JDEM-type and LSST-type surveys versus the maximum redshift (assuming, quite  unrealistically, that such surveys could actually  reach such high redshifts).

Notice from those figures that there exists
a \emph{critical} redshift $z_{cr}\sim 2-3$  beyond which the marginalised 
errors on $w_0$, $w_a$ and $\Omega_{k}$  do not improve significantly.
 The $z_{cr}$ can be understood if we consider that there is a redshift $\hat{z}$ at which the 
derivative of the angular diameter distance with respect to the
curvature changes sign~\cite{prev1}:
\begin{equation}
\left. \frac{\partial d_{A}(z=\hat{z})}{\partial \Omega_k}\right|_{\Omega_k=0}=0~,
\end{equation}
being $d_{A}(z)$ negative (positive) for redshifts below (above)
$\hat{z}$. 

For a $\Lambda$CDM cosmology with $h=0.72$ and $\omega_m=0.12$, $\hat{z} = 3.2$. Planck data will provide an exquisite measurement of $d_{A}$ 
at a redshift $ z \sim 1088$, i.e. well above $\hat{z}$. One
would thus need to measure with a good precision $d_{A}(z)$ below
$\hat{z}$ to break the dark energy-curvature degeneracy. For doing that, the maximal redshift 
covered by the survey should not be very far from $z=\hat{z}$.  For the chosen parameterization of the dark energy equation of state, going to higher
redshifts, $z \gg 2$, is unnecessary, since the errors on $w_0$, $w_a$ and
$\Omega_{k}$ will not be significantly reduced further. Of course, this conclusion depends on the assumed $w(z)$ parameterization.  

\section{Conclusions}
\label{sec:seciiii}
In this work we have assessed critically the prospects for disentangling the 
degeneracy between a non-constant dark energy equation of state parameter $w(z)$ and 
a non zero curvature $\Omega_{k}$. We have considered constraints achievable from  surveys with characteristics not too dissimilar from those of proposed  baryon
Acoustic Oscillation galaxy surveys. We have shown that despite the spectacular
resonant-like behaviour exhibited by the $w(z)$ needed to match  $H(z)$- $d_{A}(z)$ when  erroneously assuming a flat universe, the degeneracy cannot be solved in a model-independent way  from realistic observations.  Nevertheless  we  have shown that it is possible from future data to disentangle curvature from dark energy evolution if a  dark energy parameterization is assumed and  by combining BAO data with CMB constraints. Adopting the popular two-parameter $w_0, w_a$ parameterization of the  redshift evolution of the dark energy equation of state parameter we find that  measurements of both $H(z)$ and $d_{A}(z)$ up to  sufficiently high redshifts $z\sim 2$ is key to resolve the $w(z)-\Omega_k$ degeneracy.  These results are in qualitative agreement  
with e.g. Ref.~\cite{prev4}. The agreement is not quantitative but this is due  
to the fact that we consider BAO (transversal and  radial) while Ref.~\cite{prev4} concentrates on SNe. Here we have focused on
BAO surveys rather than SNe data for several reasons. The luminosity distance $d_L$ (obtained from SNe data) is an integral over $H(z)$ as it is $d_A$: adding SNe data would reduce the error-bars around $w_{d_A}$ but not around $w_H$. For parametrizations of the dark energy equation of state 
more general than the one in Eq.~(\ref{eq:eosp}), it has been shown that non-integrated quantities such as $H(z)$ allow for a better reconstruction of the dark energy dynamics than distance measurements such as $d_A$ or $d_L$, see eg. Refs. \cite{Sahni:2006pa,FernandezMartinez:2008hw}. Finally, even when the simple parametrization of Eq.~(\ref{eq:eosp}) is assumed, we have found that $d_A$ and $H(z)$ measurements at sufficiently high redshifts, around $z \sim 2$, are required to solve the $w(z)-\Omega_k$ degeneracy; currently planned SNe surveys do not reach such high redshifts. While quantitatively the results presented here will improve with the addition of SNe data, we believe we have captured qualitatively the gist of forthcoming constraints.

The conclusions presented here depend quantitatively on the parameterization chosen, and, probably to a lesser extent, on the fiducial model adopted. However, qualitatively the conclusions should hold in general; it is always possible to find a better parameterization that will yield smaller errors, but at least we have shown that with the parameterization adopted here  forthcoming surveys will lift the curvature/dark energy degeneracy (see fig 7 and 8). This conclusion, however, relies on the fact that the adopted parameterization includes the underlying model. Should there be an indication of a deviation from constant $w(z)$, the issue of the $w(z)$ parameterization will become crucial for obtaining meaningful results.

\section*{Acknowledgments}
We would like to thank E.~Masso for encouraging us to explore the degeneracy we focus on in  this paper. Our calculations made extensive use of the \emph{Beowulf cluster} (IFT-CSIC, Madrid, Spain). G.~Barenboim acknowledges support from the Spanish MEC and FEDER under Contract FPA 2008/02878, and from the Generalitat valenciana under grant, PROMETEO/2008/004. O.~Mena work is supported by a \emph{Ram\'on y Cajal} contract from MEC, Spain. O.~Mena and L.~Verde acknowledge support of MICINN grant AYA2008-03531. L. Verde acknowledges support of MICINN grant FP7-PEOPLE-2002IRG4-4-IRG\#202182. L. Verde acknowledges hospitality of CERN theory division where the last stages of this work were carried out. EFM acknowledges support by the DFG cluster of excellence ``Origin and Structure of the Universe''.  

\bibliographystyle{JHEP}

\providecommand{\href}[2]{#2}\begingroup\raggedright

\end{document}